\def\kms{km s$^{-1}$\,}

\def\hii{H{\sc ii}\,}

\def\msun{M$_\odot$\,}
\def\mjyb{mJy beam$^{-1}$}

\def\cm2{cm$^{-2}$}
\def\cm3{cm$^{-3}$}
\def\x{$\times$}

\def\ngc{\rm NGC\,3503\,}
\def\sfo{\rm SFO\,62\,}

\def\msx{\rm MSX\,}

\def\Tr{$T_R^\ast$}
\documentclass[longauth]{aa} 

\usepackage{natbib}
\usepackage{graphicx}
\usepackage{txfonts}
\begin{document}
\title{NGC\,3503 and its molecular environment}

\author{N. U. Duronea\inst{1}
          \and
           J. Vasquez\inst{1,2}
          \and C. E. Cappa \inst{1,2}
          \and M. Corti\inst{1,2}   
          \and E. M.  Arnal\inst{1,2} 
          }
\institute{Instituto Argentino de Radioastronomia, CONICET, CCT-La Plata, 
 C.C.5., 1894, Villa Elisa, Argentina   \email{duronea@iar.unlp.edu.ar}\and Facultad de Ciencias Astron\'omicas y Geof\'isicas, Universidad Nacional de La Plata,Paseo del Bosque s/n, 1900 La Plata,  Argentina} 

\date{Received 2011 August 26; accepted 2011 November 14}
 
 
\abstract
   {  }
   {We present a study of the molecular gas and interstellar dust  distribution  in the environs of the \hii\ region NGC\,3503 associated with the open cluster Pis\,17 with the aim of investigating the spatial distribution of the molecular gas linked to the nebula and achieving a better understanding of the interaction of the nebula and Pis\,17 with their molecular environment.
}
   {We based our study in $^{12}$CO(1-0) observations of a region of  $\sim$ 0\fdg 6 in size obtained with the 4-m NANTEN telescope, unpublished radio continuum data at 4800 and 8640 MHz obtained with the ATCA telescope, radio continuum data at 843 MHz obtained from SUMSS, and available IRAS, MSX, IRAC-GLIMPSE, and MIPSGAL images.
}
   {We found a molecular cloud  (Component 1) having a mean velocity of --24.7 \hbox{\kms}, compatible with the velocity of the ionized gas, which is associated with the nebula and its surroundings. Adopting a distance of 2.9 $\pm$ 0.4 kpc the total molecular mass and density yield  \hbox{(7.6 $\pm$ 2.1) $\times$ 10$^3$ \msun} and  \hbox{400 $\pm$ 240 cm$^{-3}$}, respectively.

The radio continuum data confirm the existence of an electron density gradient in NGC\,3503. The IR emission shows the presence of a PDR bordering the higher density regions of the nebula. The spatial distribution of the CO emission shows that the nebula coincides with a molecular clump, with the strongest CO emission peak  located close to the higher electron density region. The more negative velocities of the molecular gas (about --27 \hbox{\kms}), is coincident with NGC\,3503. Candidate YSOs were detected  towards the \hii region, suggesting that embedded star formation may be occurring in the neighbourhood of the nebula. The presence of a clear electron density gradient, along with the spatial distribution of the molecular gas and PAHs in the region indicates that \ngc is a blister-type \hii\ region that probably has undergone a champagne phase.     
}


{ }

\keywords{ISM: molecules, Radio Continuum: ISM, Infrared: ISM, ISM: \hii regions, ISM:individual object: \ngc, Stars: Pismis 17}

\maketitle

\section{Introduction}

  It is well established  that massive stars have a significant impact on the dynamics and energetics of the interstellar medium (ISM) surrounding them. In the classical scenario, OB stars are born  deeply buried within dense molecular clumps and emit a copious amount of far ultraviolet (FUV) radiation \hbox{($h\nu$  $>$ 13.6 eV)}. FUV photons ionize the neutral hydrogen, creating an \hii region which expands into the molecular cloud due to the pressure difference between the molecular and ionized gas.  At the interface between  the ionized and molecular gas, where photons of lower energies \hbox{(6 eV $<$    $h\nu$  $<$ 13.6 eV)} dominate, photodissociation regions (PDR) are known to exist (see \citealt{ht97} for a complete review). 

 The molecular component of \hii regions has been studied in detail by many authors \citep{kh89,rc04}. Many efforts have been devoted to observing the interaction between the \hii regions and their parental molecular clouds and some evidence has been presented showing that molecular gas near \hii regions may be kinematically disturbed by several  \hbox{\kms} \citep{ew87}. Understanding the complex interaction between massive young OB stars, \hii regions, and molecular gas is crucial to the study  of massive star formation and the impact of massive stars on their environment.

\begin{figure*}
\centering
\includegraphics[width=170mm]{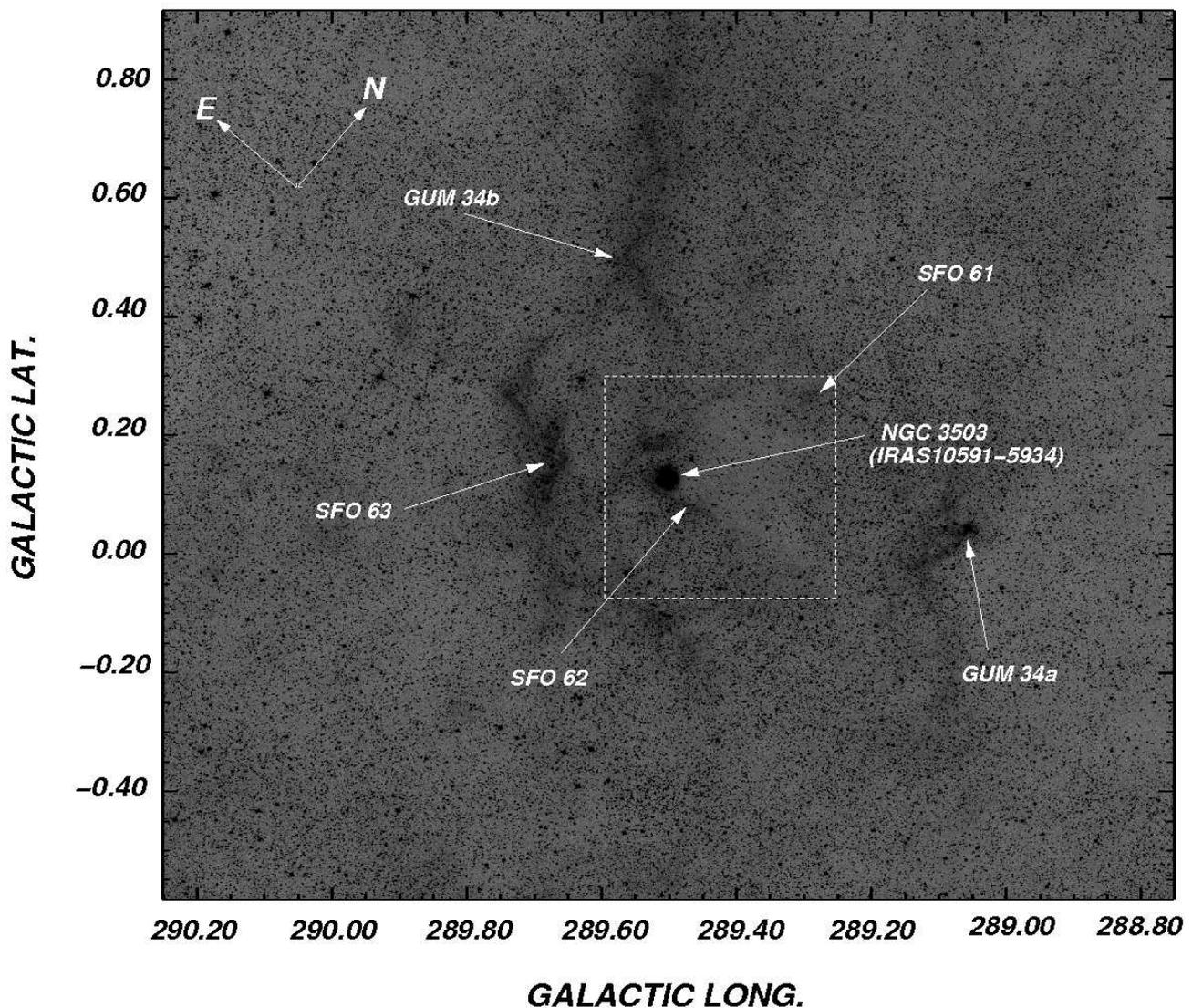}
\caption{ DSSR image of the brightest part of RCW 54. The image is  $\sim$  1\fdg5 $\times$ 1\fdg5 in size, centered at  \hbox{({\it l,b}) = (289\fdg5, 0\fdg25)}. Our region of interest is delimited by the white dashed square. }
\label{fig:rcw54}
\end{figure*}

       NGC\,3503 (=Hf 44=BBW 335)  \hbox{({\it l,b} =  289\fdg 51, +0\fdg 12)} is a bright and small ($\sim$ 3$'$ in diameter) optical emission nebula  \citep{dr88}  ionized by early B-type stars belonging to the open cluster  Pis 17 \citep{H75, pco10}   and it is seen projected onto an extended region of medium brightness in H$\alpha$ known as {\rm RCW}\,54. This last region is centered at \hbox{({\it l, b}) = (289\fdg4, -0\fdg6)}, and has dimensions \hbox{($\Delta l$ $\times$ $\Delta b$) = (3\fdg5 $\times$ 1\fdg0)} \citep{r60}. The  point source {\rm IRAS}\,10591-5934, which appears projected onto \ngc, is the infrared counterpart of the optical nebula. For the sake of clarity, we show  the main components and their relative location in Fig. {\ref{fig:rcw54}} superimposed on a DSSR image of the region of interest. The dashed white square delimits the region studied in this paper.\

In their Southern Hemisphere catalogue of bright-rimmed clouds ({\rm BRC}s) associated with {\rm IRAS} point sources, \citet{SO94} quote {\rm SFO}\,62  as related to {\rm NGC}\,3503.
BRCs are defined as isolated molecular clouds located at the edges of evolved \hii regions, and are suspected to be potential sites of star formation through the radiation-driven implosion (RDI) process \citep{so91,U09}.   Making use of  the Australian Telescope Compact Array ({\rm ATCA}),  \citet{T04} carried out radio continuum observations towards all {\rm BRC}s catalogued by \citet{SO94}. These authors classified {\rm SFO}\,62 as a broken-rimmed cloud associated with an evolved stellar cluster that is about to disrupt its natal molecular cloud. Latter on, \citet{U09} claimed that SFO62 was incorrectly classified as a BRC, and that the RDI process is not working in this region.

Using the [S{\sc ii}] $\lambda$6716/$\lambda$6731 line ratio of \ngc, \citet{c00} detected a significant electron density dependence on position, which according to the authors, could be interpreted as a radial gradient. The authors claimed that \ngc may be a  candidate of showing a ``champagne flow''. In their H$\alpha$ survey of the Milky Way towards \hbox{{\it l} = 290$^{\circ}$}, \citet{g00} reported a radial velocity\footnote{Radial velocities are referred to the Local Standard of Rest ({\rm LSR})} of  \hbox{--21 \hbox{\kms}} towards  \ngc, and  proposed that \ngc is linked to a complex of H{\sc ii} regions placed at a distance of about 2.7 kpc, with radial velocities of about $-$25 \hbox{\kms}. Published distance determinations of {\rm NGC}\,3503 vary between a minimum of 2.6 kpc \citep{H75} and a maximum of 4.2 kpc \citep{mv75}. In this paper we shall adopt a distance of \hbox{2.9 $\pm$ 0.4 kpc} \citep{pco10}.

 {\rm NANTEN} $^{13}$CO (J=1$\rightarrow$0) (HPBW = 2.7$'$) line observations were carried out by \citet{Y99}   towards 23 H{\sc ii} regions associated with 43 {\rm BRC}s catalogued by \citet{SO94}, with a grid spacing of $\sim$ 4$'$ $\times$ 4$'$. The aim of that work was to investigate statistically  the dynamical effects of H{\sc ii} regions on their associated molecular clouds and star formation.   A square region of $\sim$ 0\fdg4 in size towards {\rm NGC}\,3503 was observed  and a single and slightly extended  6$'$ diameter  molecular cloud   having a peak radial velocity of --24.9 \hbox{\kms} was found. Adopting a distance of 2.9 kpc, this molecular structure has a linear radious of 2 pc and a total mass of $\sim$ 500 \msun. The authors rejected this cloud from the analysis sample, since it was detected at only one position.  $^{12}$CO, $^{13}$CO, and $^{18}$CO (J=1$\rightarrow$0) {\rm MOPRA} position-switched observations were carried out  by \citet{U09}. The {\rm ON} position was centred on the position of the {\rm IRAS} source associated with {\rm NGC}\,3503. The least abundant of the isotopes was not detected, whilst the emission lines of the other two isotopes show a double-peaked profile. From a Gaussian fitting, their peak radial velocities are  $\sim$ +20 \hbox{\kms} and --25.6 \hbox{\kms}. 

Molecular line observations provide an invaluable support in pursuing  a better understanding of the interaction of H{\sc ii} regions with their surroundings. Previous observations of \citet{Y99} and \citet{U09}, although providing important information about the molecular gas associated with \ngc, do not offer a complete picture of the molecular environment of the nebula. Bearing this in mind, the goals of this paper are twofold, namely: {\it a)} to map the spatial distribution of the  molecular gas associated with {\rm NGC}\,3503 and to study its physical characteristics, and  {\it b)} to achieve a better understanding of the interaction of {\rm NGC}\,3503 and Pis 17 with their molecular environment. These aims were addressed by analyzing new $^{12}$CO (J=1$\rightarrow$0) data gathered by using the {\rm NANTEN} telescope to observe a square region \hbox{$\sim$ 0\fdg7} in size, centred on \hbox{({\it l,b}) = (289\fdg36, +0\fdg02)}.  Molecular observations were combined with  unpublished radio continuum data, and optical,  mid, and \hbox{far-infrared} archival data. 

The structure of the paper is the following. In Sect. 2,  the {\rm CO} and radio continuum observations are briefly described along with a short mention to the archival data used in this work. The main observational results of the molecular, ionized and dust emission are detailed in Sect. 3, while both the comparison among these interstellar phases and the action of the stellar ionizing sources of NGC\,3503 over its interstellar environs, as well as the presence of star formation activity are exposed in Sect. 4. Finally, a possible scenario is put forward to explain the stellar and interstellar interactions. A summary is presented in Sect. 5.


\section{Observations and data reductions}

The databases used in this work are:

\begin{enumerate}
\item Intermediate angular resolution, medium sensitivity, and
high-velocity resolution {\rm $^{12}$CO (J=1$\rightarrow$0)} data
 obtained with the 4-m {\rm NANTEN} millimeter-wave telescope of Nagoya University. At the time the authors carried out the observations, April 2001, this telescope was
 installed at Las Campanas Observatory, 
Chile. The half-power beamwidth and the system temperature, including the
atmospheric contribution towards the zenith, were  2\farcm7
($\sim$ 2.3 pc at 2.9 kpc) and $\sim$ 220\,K
 (SSB) at 115 GHz, respectively. The data were gathered using the position switching mode. Observations of points devoid of {\rm CO} emission were interspersed among the
 program positions. The coordinates of these points were retrieved from a
database that was kindly made available to us by the {\rm NANTEN} staff. The
 spectrometer used was an acusto-optical with 2048 
channels providing a velocity resolution of $\sim$ 0.055  kms$^{-1}$. For
intensity calibrations, a room-temperature chopper wheel was employed \citep{pb73}. 
An absolute  intensity calibration 
 \citep{uh76,ku81} was achieved by observing
Orion {\rm KL} \hbox{(RA(1950.0) = 5$^h$\,32$^m$\,47\fs0},\hbox{Dec (1950.0) = $-$5$^\circ$\,24\arcmin\,21\arcsec )}, and $\rho$ Oph East  \hbox{(RA(1950.0) = 16$^h$ \,29$^m$ \,20\fs9}, \hbox{Dec (1950.0) = $-$24$^\circ$\,22\arcmin\,13\arcsec )}. The absolute radiation
 temperature, $T_R^\ast$, of
Orion {\rm KL} and $\rho$ Oph East, as observed by the {\rm NANTEN} 
radiotelescope were assumed to be 65 K and 15 K, respectively \citep{myom01}. 
The {\rm CO} observations  covered a region ($\bigtriangleup${\it l}
 $\times$ $\bigtriangleup${\it b}) of 35\farcm1 $\times$ 35\farcm1 centred at 
 ({\it l, b}) = {(289\fdg36, +0\fdg02 )}     and the observed grid consists of points located  every 
one beam apart. A total of 169 positions were observed. Typically, the integration time per point was 16s resulting in an rms noise of $\sim$ 0.3 K. A second order degree polynomial was substracted from the observations to account for instrumental baseline effects. The spectra were reduced using CLASS software (GILDAS working group).

\item Unpublished radio continuum data at 4800 and 8640 MHz which were kindly provided by \hbox{J. S. Urquhart} and \hbox{M. A. Thompson}. The images were obtained in March 2005 with the Australia Telescope Compact Array (ATCA) with synthesized beams and rms noises of \hbox{23\farcs 55 $\times$ 18\farcs 62} and \hbox{0.82 \mjyb} at 4800 MHz and \hbox{14\farcs 73 $\times$ 11\farcs 74} and \hbox{0.56 \mjyb} at 8640 MHz. Radio continuum archival images were also obtained from the Sydney University Molonglo Sky Survey (SUMSS)\footnote{{\it http://www.astrop.physics.usyd.edu.au/cgi-bin/postage.pl}} \citep{b99} at 843 MHz, with angular resolution of 45$''$$\times$45$''$cosec($\delta$).

\item Infrared data retrieved from the  {\it Midcourse Space Experiment (MSX)}\footnote{{\it http://irsa.ipac.caltech.edu/Missions/msx.html}} \citep{p01}, high resolution IRAS images (HIRES)\footnote{http://irsa.ipac.caltech.edu/applications/IRAS/IGA/} at 60 and 100 $\mu$m   \citep{fa94}, Spitzer images at 8.0  and 4.5 $\mu$m from the Galactic Legacy Infrared Mid-Plane Survey Extraordinaire (GLIMPSE)\footnote{{\it http://sha.ipac.caltech.edu/applications/Spitzer/SHA//}} \citep{b03}, and Multiband Imaging Photometer for $Spitzer$ (MIPS) images at 24 and 70 $\mu$m from the MIPS Inner  Galactic Plane Survey (MIPSGAL)\footnote{{\it http://sha.ipac.caltech.edu/applications/Spitzer/SHA//}} \hbox{\citep{ca05}}. 

\item Optical data retrieved from the 2nd Digitized Sky Survey  (red plate)\footnote{{\it http://skyview.gsfc.nasa.gov/cgi-bin/query.pl}}  \citep{mcl00}.

\end{enumerate}

\section{Results and analysis}

\subsection{Carbon Monoxide}

\subsubsection{Individual velocity components}
		   
\begin{figure*}
\centering
\includegraphics[width=430pt]{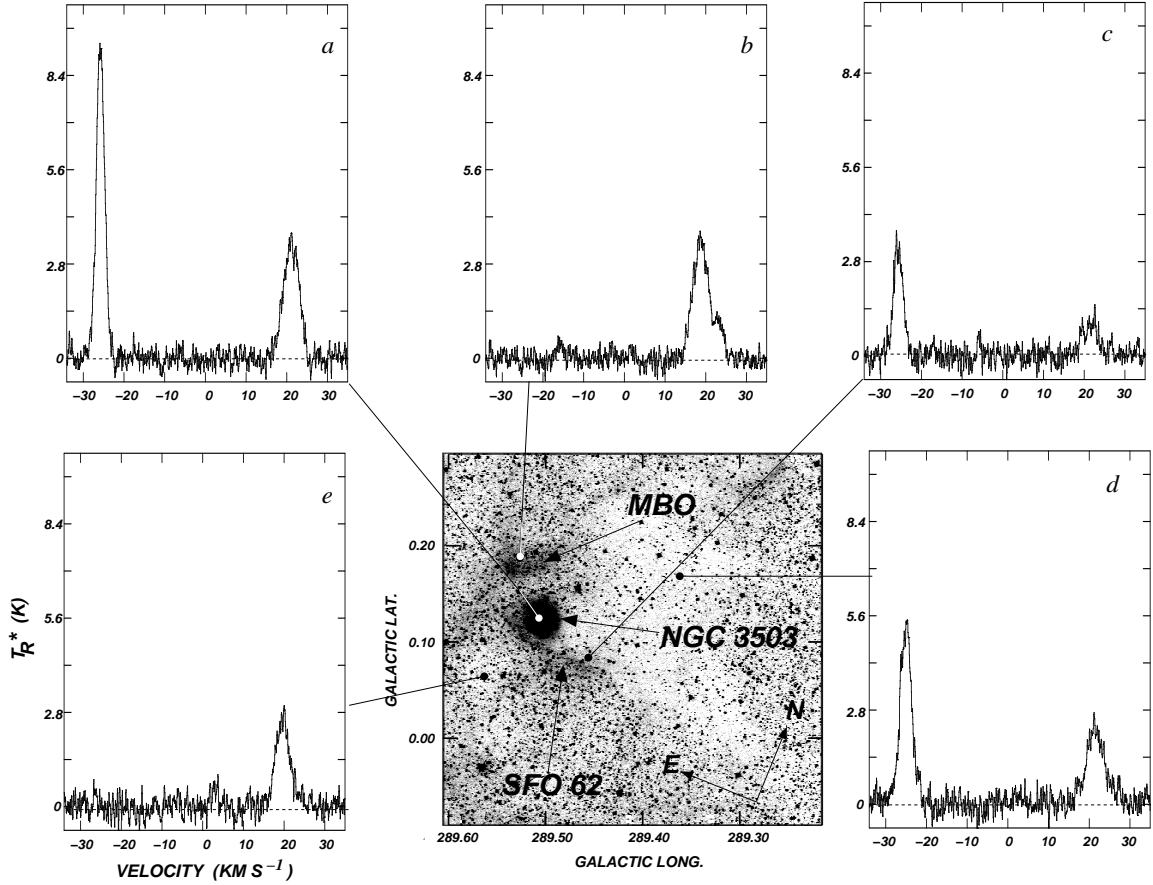}
\caption{CO emission profiles towards five positions around  NGC 3503. The CO profiles are averaged over a square area $\sim$ 3$'$ in size, centered on the black/white dots drawn on the DSSR image (center). The intensities  are given as absolute radiation temperature \Tr. }
\label{fig:perfiles}
\end{figure*}

In order to illustrate in broad terms the molecular structures detected towards the region under study, a series of {\rm CO} profiles are displayed in Fig.~\ref{fig:perfiles}.

Profiles {\it a)} and {\it c)} show the {\rm CO} emission along the line of sight to \ngc and the bright edge of \sfo, respectively. In both spectra the bulk of the molecular emission is detected between \hbox{--30} and \hbox{--20} \hbox{\kms}, and +15 to +25 \hbox{\kms}, in good agreement with previous results by \citet{U09}.

 The medium brightness optical region northeast of \ngc\ (from here onwards MBO) though also depicts a double peak structure (see profile {\it b}), displays some small scale structure in the feature peaking at $\sim$ +20 \hbox{\kms}, while the most negative CO feature is detected at $\sim$ --18 \hbox{\kms}. 

The molecular gas along regions of high optical absorption is shown in profiles {\it d)} and {\it e)}. The former resembles  the CO spectrum observed towards \ngc, and is characteristic of the molecular emission in the region of high optical absorption seen northwest of \ngc. The roundish patch of high absorption seen at \hbox{{\it (l,b)} = (289\fdg55,+0\fdg07)} only shows the CO peak at positive velocities. 

\begin{figure*}
\centering
\includegraphics[width=378pt]{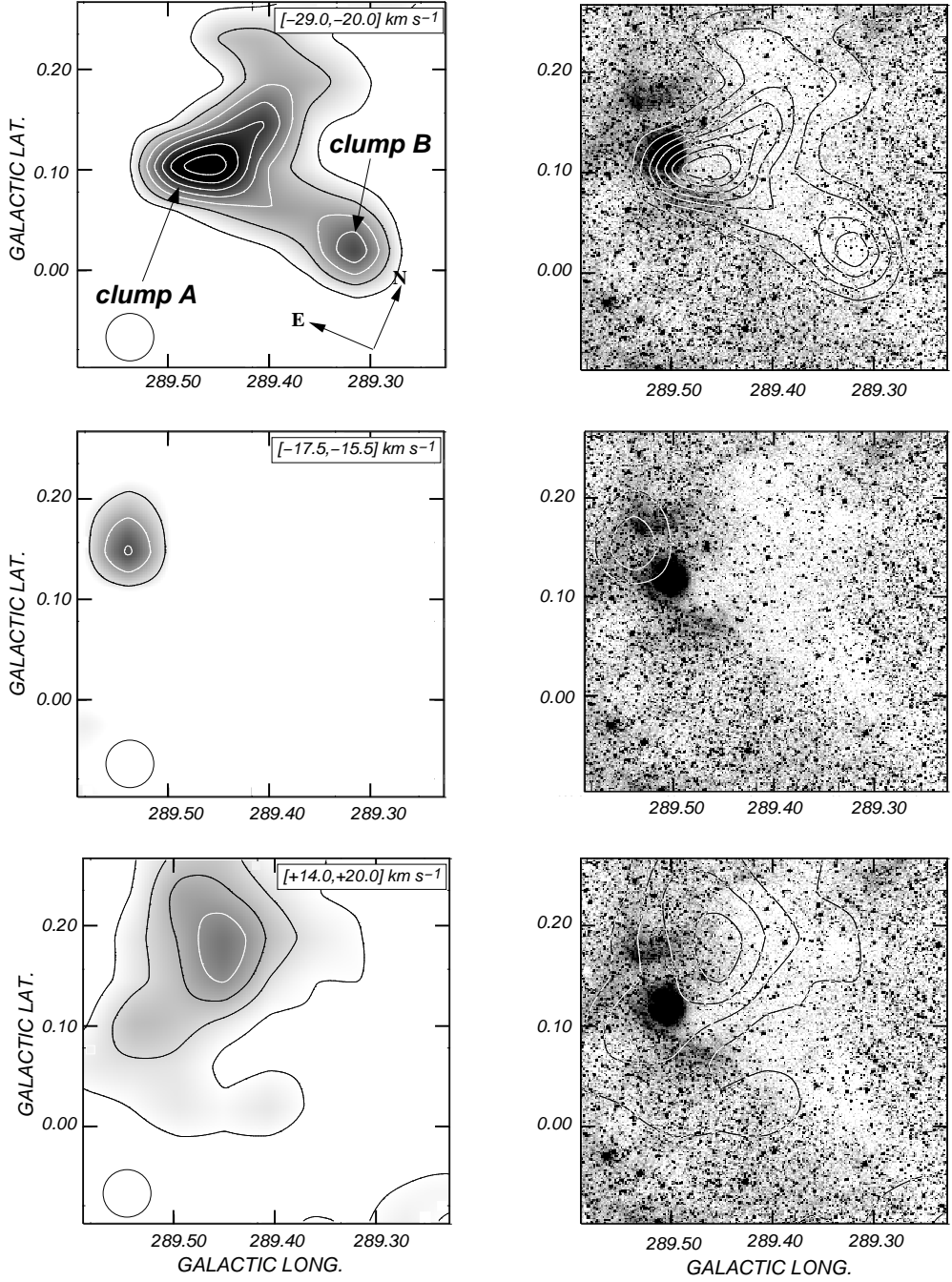}
\caption{{\it Upper panels} (Left):  Averaged \Tr in the velocity range $\sim$ --29 to -- 20 \hbox{\kms} (Component 1). Contour levels start at 0.42 K (23 rms)  and the contour spacing is 0.6 K. The beam size of the CO observations is shown by a  circle  in the lower left corner. (Right): Overlay of the  \Tr values in the same velocity interval (contours), superimposed on the DSSR image (greyscale). {\it Middle panels:} Averaged \Tr   in the velocity range $\sim$ --17.5 to --15.5 \hbox{\kms} (Component 2). Contour levels start at 0.7 K (23 rms) and the contour spacing is 1 K. {\it Lower panels:}  Averaged \Tr  in the velocity range $\sim$ +14 to +24 \hbox{\kms} (Component 3). Contour levels start at 0.35 K (23 rms) and  the contour spacing is 0.45 K. In all cases, the molecular emission grayscale goes from 0.35 K to 3.9 K. The orientation of the equatorial system is indicated in the top left panel.}
\label{fig:comp12y3}
\end{figure*}

The spatial distribution of the molecular gas observed in the three velocity intervals mentioned above is shown in the left side panels of Fig. \ref{fig:comp12y3}. In order of increasing radial  velocity, the  CO components will be referred to as Component 1 (peaking at \hbox{$\sim$ --25 \kms}), Component 2 (peaking at \hbox{$\sim$  -16 \kms}), and Component 3 (peaking at \hbox{$\sim$ +20 \kms}). To facilitate the comparison between the spatial distribution of molecular and ionised gas, the right panels of Fig. \ref{fig:comp12y3} show the same molecular contours as the left panels superimposed to the DSSR  image in grayscale. The difference in spatial distribution among the three components is readily appreciated. The molecular gas in the velocity interval --29 to --20 \hbox{\kms}\ (Component 1) shows two well developed concentrations, whose emission peaks are located at \hbox{{\it (l,b)} = (289\fdg47,+0\fdg12)} (clump A) and \hbox{{\it (l,b)} = (289\fdg32\arcmin,+0\fdg03)} (clump B). Component 1 has a very good morphological correspondence with the high optical obscuration region seen north and northwest of \ngc. 

Component 2 peaks at \hbox{{\it (l,b)} = (289\fdg55, +0\fdg17)} near both \ngc and MBO, and is projected onto a region without apreciable optical absorption. Finally, Component 3, peaking at \hbox{{\it (l,b)} = (289\fdg45, $-$0\fdg18)}, does not show a clear morphological correspondence either with ionized regions or areas displaying strong absorption.

For the three molecular components, a mean radial velocity ($\bar{V}$) weighted by line temperature, was derived by means of
\begin{eqnarray}
\qquad \qquad     \bar{V} \ =\  \frac{\sum_{i}\  T_{Peak_i}\ \times\  V_{Peak_i}}{\sum_{i}\  T_{Peak_i}}
 \label{eq:masaH2a}
\end{eqnarray}
where $T_{\rm Peak_i}$ and $V_{\rm Peak_i}$ are the peak  \Tr temperature and the peak radial velocity of the $i$-spectrum observed within the 3 rms contour line defining the outer border. Mean radial velocities of  \hbox{--24.7 \kms}, \hbox{--16.6  \kms}, and \hbox{+19.5 \kms} were obtained for Component 1, Component 2, and Component 3, respectively.

Based on $\bar{V}$, we tried to estimate the kinematical distance  of Component 1.  The analytical fit to the rotation curve by \citet{bb93} along {\it l} = 289\fdg5 shows  that radial velocities more negative than --15.5 \hbox{\kms}\ are forbidden for  this galactic longitude. However, the ionised gas along this galactic longitude exhibits velocity departures  more negative than $-$7 \hbox{\kms}  relative to the circular rotation model, as the cases of the \hii\ regions  Gum\,37 \hbox{{\it (l,b)} = (290\fdg65, +0\fdg26)} \hbox{($d$ = 2.7 kpc)}  and Gum\,38a  \hbox{{\it (l,b)} = (291\fdg28, +0\fdg71)} \hbox{($d$ = 2.8  kpc)}, whose radial velocities are in the range $-$25 to $-$28 \hbox{\kms} and $-$25 to $-$ 29 \hbox{\kms}, respectively \citep{g00}.   Bearing in mind that close to the tangential point along {\it l} = 290\fdg5 the radial velocity gradient is very small and that the \hii\ regions located at distances between 2.7 and 2.8 kpc exhibit forbidden radial velocities similar to those found for Component 1, we adopt for the latter a distance of 2.9 $\pm$ 0.4 kpc \citep{pco10}, which is also the kinematical distance corresponding to the tangential point.

 Component 1 shows an excellent morphological resemblance with a region of high optical absorption next to \ngc (see Fig. {\ref{fig:comp12y3}}) and its mean velocity is in good agreement with the velocity of the H$\alpha$ line towards the nebula ($-$21 \hbox{\kms}; \citealt{g00}). It is also worth noting that the velocity interval of Component 1 is in excellent agreement with the velocity of SFO\,62 found by \citet{SO94} \hbox{(--28 $\leq\ v_{LSR}\ \leq$ --20 \hbox{\kms}}, at $^{12}$CO),  by \citet{Y99} (--24.9 \kms at $^{13}$CO),  and by \citet{U09} ($-$25.6 and $-$25.7 \hbox{\kms}, at $^{12}$CO and $^{13}$CO, respectively). Based on above, we conclude that Component 1 is associated with \ngc and its environs. The bright rim  of \sfo\ clearly follows the southernmost border of clump A, which  very likely indicates that the molecular gas in the southern border of this clump is being ionized, originating the bright optical rim.

  On the other hand, considering that the mean radial velocity of Component 2 (--16.5 \hbox{\kms}) is close to the radial velocity of the tangential point at {\it l} = 289\fdg5 (--15.5 \hbox{\kms}), we suggest that Component 2  may be located in the neighborhood of the tangential point, close to \hbox{Component 1}. From the mean radial velocity of Component 3, a kinematic distance of $\sim$ 8 kpc is determined, indicating that Component 3 is unrelated to \ngc. Very likely, Component 3 is associated with the complex of H{\sc ii} regions at a velocity of $\sim$ +20 \hbox{\kms}\ reported by \citet{g00}.

 A direct comparison of Component 1 with the molecular cloud detected by \hbox{\citet{Y99}}  (see Fig. 2u from that work), shows that the angular size of Component 1 is about a factor 3 - 4 greater than the latter. Clearly, only the densest part of the molecular cloud associated with \ngc (clump A) was detected in the $^{13}$CO observations of \hbox{\citet{Y99}}. From here onwards, the analysis of the molecular gas associated with \ngc will focus on Component 1.

\subsubsection{Kinematics and excitation conditions}

\begin{figure}
\centering
\includegraphics[width=240pt]{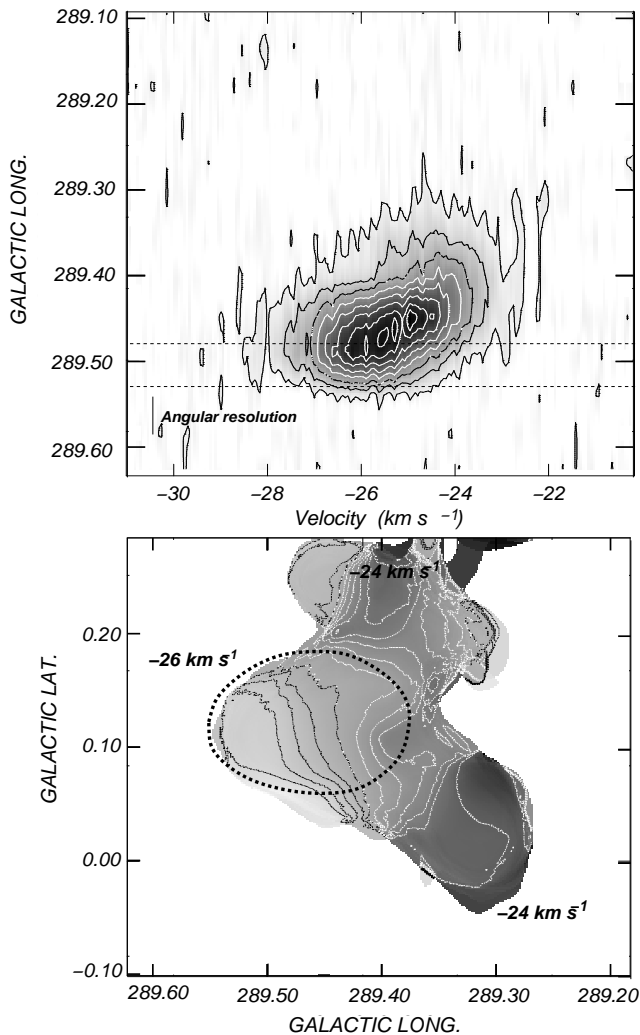}
\caption{{\it Upper panel}: Velocity-Galactic Longitude map obtained in a strip along {\it b} = +0\fdg11   (corresponding to clump A)  showing \Tr. Contour levels start at 0.7 K and the contour spacing is 0.7 K. The dotted lines indicate the location of \ngc. {\it Lower panel}:   Mean velocity map of Component 1. Contours levels go from $-26$  to $-24$ \hbox{\hbox{\kms}}, with interval of 0.2 \hbox{\hbox{\kms}}.  The dotted ellipse depicts approximately the region of clump A.    }
\label{fig:pos-vel}
\end{figure}

The kinematics of Component 1 was studied by using position-velocity maps across selected strips. The map obtained along \hbox{{\it b} = +0\fdg12}  (corresponding to clump A) is shown in the upper panel of Fig.~\ref{fig:pos-vel}.  A noticeable velocity gradient is observed at velocities from about $-23$ \hbox{\kms} to about $-28$ \hbox{\kms}.

Additional support to the existence of this gradient can be obtained through a moment analysis.  Due to the large angular dimensions of Component 1  and the spatial sampling of NANTEN observations, 28 independient CO profiles were observed towards the region enclosed by the 0.42 K-contour line in Fig.~\ref{fig:comp12y3}. Having these spectra a high signal-to-noise ratio \hbox{(S/N $>$ 10)}, they are well suited to study in some detail possible variations of the CO profiles across Component 1. With this aim, we used the AIPS package to calculate the first three moments (integrated area, temperature weighted mean radial velocity, and velocity dispersion) of the CO profiles. In the lower  panel of Fig. ~\ref{fig:pos-vel}, the  temperature weighted mean radial velocity  distribution of CO is shown. A  clear velocity gradient is observed along Component 1, and particulary  in the region of clump A (depicted by the dotted ellipse), with mean radial velocities more negative toward  the position of \ngc. A velocity shift of about $\sim$ 1.4 \hbox{\kms} is observed across  the region of clump A, which at a distance of 2.9 kpc translates to a gradient $\omega$ $\approx$ 0.15  \hbox{\kms} pc$^{-1}$.  This panel also shows that  the CO emission corresponding to clump B does not show a significant velocity gradient.

In order to offer a complete picture of the kinematical properties of clump A, we show in Fig.~\ref{fig:mosaico1} the spatial distribution of the CO emission within the velocity range from --27.4 to --23.4 \hbox{\kms}. Every image depicts mean \Tr-values  (in contours) over a velocity interval of \hbox{1 \hbox{\kms}} superimposed on the 8.13 $\mu$m \msx\ emission (in greyscale). In the velocity range from --27.4 to --26.4 \hbox{\kms} the molecular emission arising from clump A is slightly displaced from the brightest MSX emission located at  {\it (l,b)} $\approx$ (289\fdg5, +0\fdg11). As we move towards more positive velocities, the maximum of the CO emission gradually shifts westwards.

\begin{figure}
\centering
\includegraphics[width=255pt]{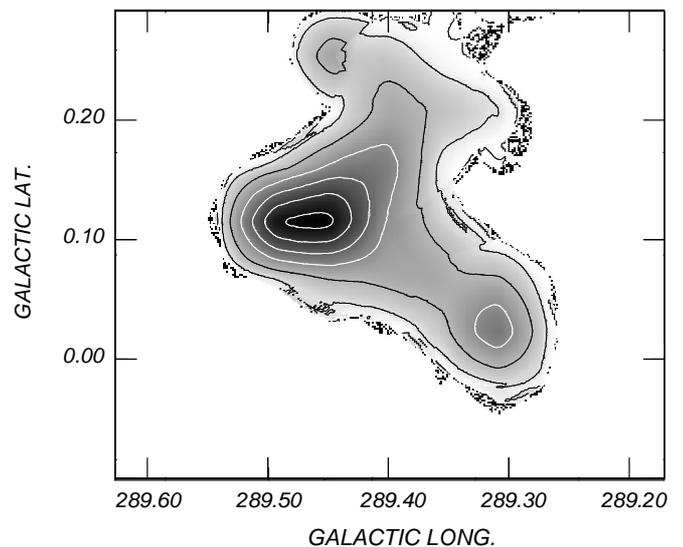}
\caption{ Spatial distribution of $T_{\rm exc}$ for Component 1. Contours leves are 5.7, 7.9, 10.2, 12.4, 14.5, and 16.6 K. }
\label{fig:momento}
\end{figure}

\begin{figure*}
\centering
\includegraphics[width=175mm]{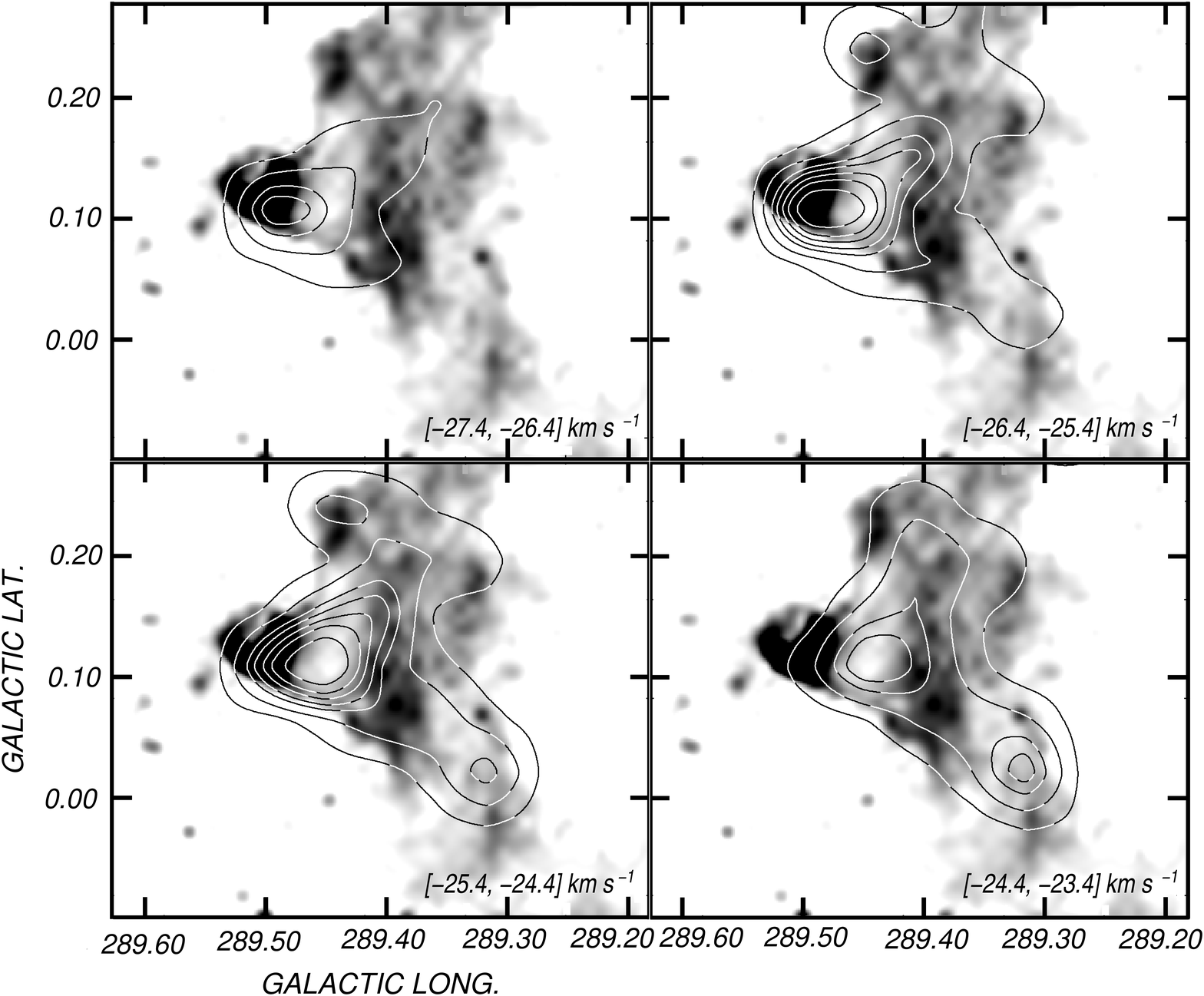}
\caption{Overlay of  mean \Tr-values (contours) in the velocity range from $-$27.4 to $-$23.4 km s$^{-1}$ and the MSX band A emission (greyscale). Every image represents the CO emission distribution averaged in a velocity interval of 1 km s$^{-1}$. The velocity interval is indicated in the bottom right corner of each image. The lowest temperature contour is 0.84 K ($\sim$ 12 rms). The contour spacing temperature is 1.4 K.}
\label{fig:mosaico1}
\end{figure*}

Following \citet{U09}, we analyze the excitation temperature ($T_{\rm exc}$) obtained from the CO data to probe the surface conditions of Component 1. 
The excitation temperature of the $^{12}$CO line can be obtained considering that $^{12}$CO is optically thick ($\tau_{\nu}$ $>>$ 1). Then, the peak temperature of this line is given by
\begin{equation}\label{eq:tpeak}
  T_{peak}\ (^{12}CO)\ =\ J_{\nu}(T_{exc})\ -\ J_{\nu}(T_{bg})
\end{equation} 
\citep{di78} where $J_{\nu}$ is the Planck function at a frequency $\nu$. Assuming gaussian profiles for the $^{12}$CO line, combining the order zero moment map (i.e., integrated area) with the order two moment map (i.e., velocity dispersion), and using Eq.~ \ref{eq:tpeak} we obtain the $T_{\rm exc}$ distribution map \hbox{(Fig. \ref{fig:momento})}. As expected, the $T_{\rm exc}$ distribution is quite similar to the CO emission distribution of Component 1, reaching a  maximum of $\sim$ 17.7 K at \hbox{({\it l,b}) $\approx$ (289\fdg45, +0\fdg11)}.       It is worth noting that the obtained value of $T_{exc}$ towards the center of clump A is lower than the obtained by \citet{U09} towards \hbox{{\it (l,b)} $\approx$ (289\fdg5, +0\fdg11)}     \hbox{($T_{\rm exc}$ = 23.9 K)}. This difference may be explained in terms of a beam smearing of our NANTEN data, which implies that the values of $T_{\rm exc}$ shown in Fig. \ref{fig:momento} must be considered as lower limits.

\subsubsection{Masses and densities}

   The  mass of the molecular gas  associated with \ngc can be derived  making use of  the empirical relationship between the molecular hydrogen column density, $N(\rm H_2)$, and the integrated molecular emission, \hbox{$I_{{\rm ^{12}CO}}$ ($\equiv\int\ T^*_{R}   \ d{\rm v}$)}. The conversion between $I_{{\rm ^{12}CO}}$  and $N(\rm H_2$) is given by the equation 
\begin{equation}\label{eq:cero}
 \quad   N({\rm H_2})\ =\ (1.9\ \pm\ 0.3)\ \times\ 10^{20}\ I_{{ ^{12}CO}} \ \ \ \quad  ({\rm cm}^{-2})
\end{equation}
  \citep{ d96,sm96}. The total  molecular mass $M_{\rm tot}$, was calculated through
\begin{equation}\label{eq:uno}
  \quad  M_{\rm tot}\ =\  (m_{sun})^{-1}\  \mu\ m_H\ \sum\ \Omega\ N({\rm H_2})\ d^2 \qquad  \quad  \ \quad ({\rm M}_{\odot})
\end{equation}
where  $m_{sun}$ is the solar mass ($\sim$ 2 $\times$ 10$^{33}$ g),    $\mu$ is the mean molecular weight, assumed to be equal to 2.8 after allowance of a relative helium abundance of 25\% by mass \citep{Y99,ya06},  $m_{H}$ is the 
hydrogen atom mass   ($\sim$ 1.67 $\times$ 10$^{-24}$ g), $\Omega$ is the solid angle subtended by the CO feature  in ster, $d$ is the  distance, expressed in cm, and  $M_{\rm tot}$ is given in solar masses. For Component 1, we obtain a mean column density of $N(\rm H_2)$ = (1.9 $\pm$ 0.3) $\times$ 10$^{21}$ cm$^{-2}$,  and a total molecular mass of \hbox{$M_{\rm tot}$ = (7.6 $\pm$ 2.1) $\times$ 10$^3$ \msun}. 
The uncertainty in $M_{tot}$ ($\sim$ 28 $\%$) stems from the distance uncertainty and from the error  quoted for the coefficient in Eq.~\ref{eq:cero}. Areas having  $T^*_{R}$ $\geq$ 0.42 K were taken into account. The difference with the value derived by \citet{Y99}, (500 \msun),  is due to both the subsampled observations of these authors and the fact that $^{13}$CO line is a better tracer of high density regions.

   The mean volume density ($n_{\rm H_2}$) of Component 1  can be derived from the ratio of its molecular mass and its volume considering that the volume of Component 1 is the result of the addition  of an ellipsoid with mayor and minor axis of 7\farcm 5 and 4\farcm 5, respectively, centered at \hbox{{\it (l,b)} $\approx$ (289\fdg42, +0\fdg15)} (which includes clump A), and a sphere of 4\farcm 5 in radius, centered at \hbox{{\it (l,b)} $\approx$ (289\fdg33, +0\fdg04)} (which includes clump B). We derived $n_{\rm H_2}$ $\approx$ 400 cm$^{-3}$. Taking into account the mass and volume uncertainties,  we derive a  conservative density  error  of about \hbox{60 $\%$} \hbox{($\sim$ 240 cm$^{-3}$).}

\subsection{Radio continuum emission}

\begin{figure*}
\centering
\includegraphics[width=420pt]{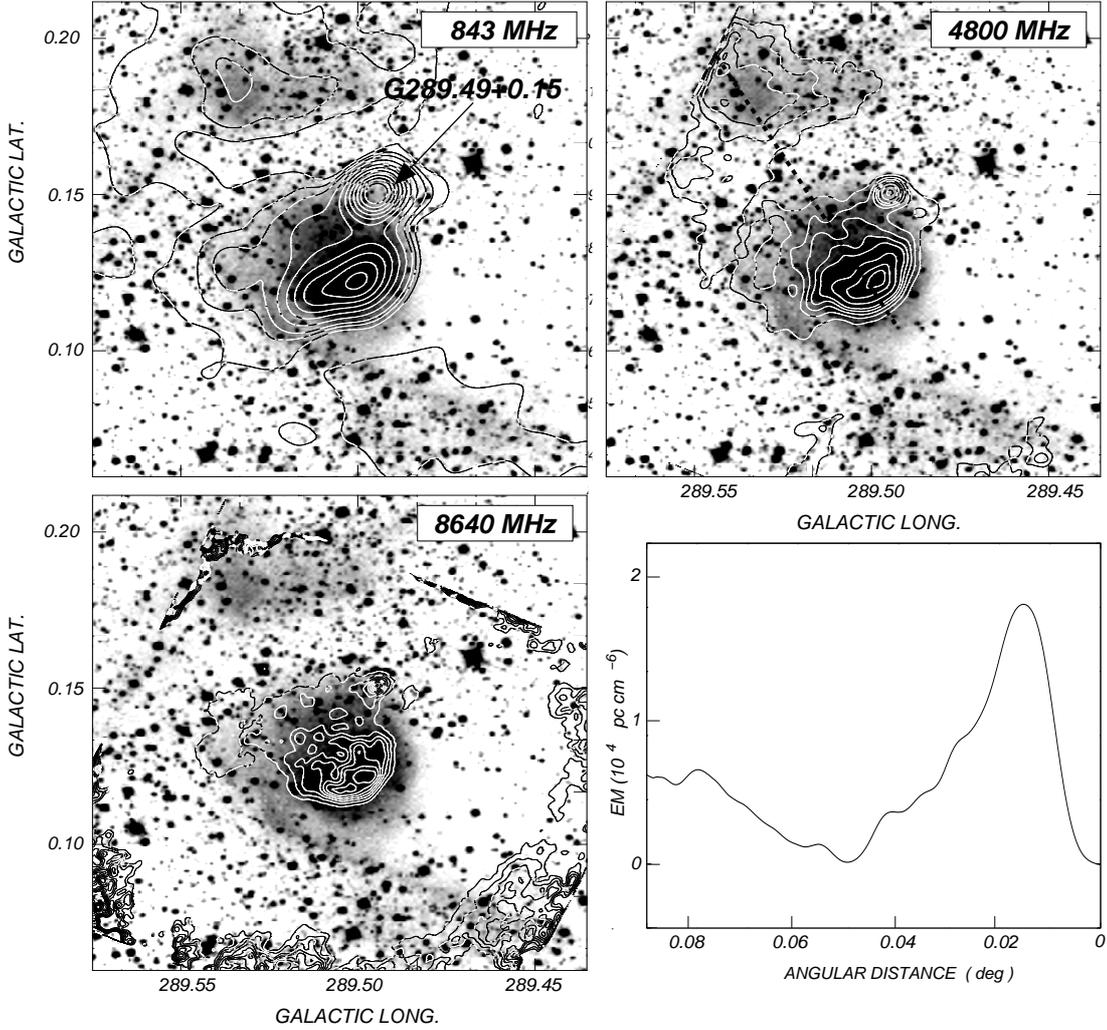}
\caption{{\it   Upper left panel:}  Radio continuum image at 843 MHz (contours)  superimposed to the DSSR  image (grayscale). Contours levels go from 4 \mjyb  \hbox{($\sim$ 3 rms)}  to 12 \mjyb   in steps of  4 \mjyb, and from 20 \mjyb in steps of 10 \mjyb. {\it   Upper right panel:}  Radio continuum image at 4800 MHz (contours) superimposed to the DSSR image (grayscale). Contours levels go from 2.4 \mjyb \hbox{($\sim$ 3 rms)}  to  10.4 \mjyb   in steps of  2 \mjyb, and from 10.4  \mjyb in steps of 4 \mjyb. The symmetry axis of the nebula is depicted by the dotted line. {\it   Lower left panel:}  Radio continuum image at 8640 MHz (contours) superimposed to the DSSR image (grayscale). Contours levels go from 15 \mjyb \hbox{($\sim$ 3 rms)}   in steps of  1 \mjyb. {\it   Lower right panel:} Emission measure profile obtained from the 4800 MHz radio continuum image along the symmetry axis of \ngc.}  
\label{fig:843}
\end{figure*}

 Figure {\ref{fig:843}}  shows the radio continuum images at 843, 4800, and 8640 MHz (in contours) superimposed on the DSSR image (in grayscale) in a region of $\sim$ 9$'$ $\times$ 9$'$ centred on \ngc.

   The continuum images show an extended source coincident with \ngc. The maxima at the three frequencies coincide with the brightest optical emission region. The source, which has a good morphological correspondence with the optical emission,   exhibits a cometary shape with its symmetry axis indicated by the dotted line in the image at 4800 MHz  (Fig. {\ref{fig:843}, top right panel)}.  The  fainter emission area   detected at 843 and 4800 MHz to the northeast of \ngc coincides with MBO and appears to be linked to the emission of the nebula at \hbox{({\it l,b}) $\approx$ (289\fdg53, +0\fdg13)}.  The emission region  detected at  \hbox{{\it(l,b)} $\approx$ (289\fdg48, +0\fdg08)} at 843 MHz depicts some morphological correspondence with the bright rim of SFO 62. The point-like source placed at  \hbox{({\it l,b}) $\approx$ (289\fdg49,+0\fdg15)} without  optical emission counterpart is labelled G289.49+0.15 in Fig. {\ref{fig:843}}.  From its flux densities \hbox{$F_{843}$ = 100 mJy}, \hbox{$F_{4800}$ =  17.1 mJy}, and  \hbox{$F_{8640}$ = 11.5 mJy}, we derived  a spectral index $\alpha$ = --0.92 $\pm$ 0.03   \hbox{(S$_{\nu}$ $\propto$ $\nu^{\alpha}$)}, which suggests that G289.49+0.15 is a non-thermal extragalactic source. We substracted the emission of this  source in order  to calculate radio continuum flux densities of \ngc. The results are given in Table \ref{table:flujos-cont}. The spectral index based on these estimates is \hbox{$\alpha$ = $-$0.12 $\pm$ 0.03}, typical for the thermal free-free radio continuum emission of an H{\sc ii} region.

 \begin{table}
\caption{Radio continuum flux densities measurements of NGC\,3503}
\begin{center}
\begin{tabular}{ l c c c}
\hline
Frequency & 843 MHz  & 4800 MHz & 8640 MHz \\

\hline\hline 
&&&\\
 Flux density (mJy) &  320 $\pm$ 30   &  270 $\pm$ 50   &  240 $\pm$ 40    \\
&&&\\
\hline
\label{table:flujos-cont}
\end{tabular}
\end{center}
\end{table}

We  estimate  the emission measure $EM$ $\left(= \int n_e^2\ dl \right)$ along the symmetry axis of the nebula using the image at 4800 MHz.   The brightness temperature ($T_b$)  is related to the optical depth ($\tau$) by
\begin{eqnarray}
\quad    T_b(\nu) = T_e \times (1-e^{-\tau})
\label{formula:tb}
\end{eqnarray}
where $T_e$ = 8100 $\pm$ 700 K  \citep{q06} is the electron temperature and the optical depth $\tau$  is given by
\begin{eqnarray}
\quad     \tau = 8.235\ \times\ 10^{-2}\ T_e^{-1.35}\ \nu^{-2.1}\ EM
\label{formula:tau}
\end{eqnarray}
In this last expresion, $\nu$ is given in GHz, and $EM$  in pc cm$^{-6}$. The $EM$ profile is shown in the lower right panel of Fig. {\ref{fig:843}}. As expected, two maxima are observed, one at the brightest section of \ngc and the other coincident with  MBO, indicating that the electron density is higher towards  these regions than in the sorroundings. The $EM$ profile of \ngc is consistent with the cometary shape morphology of the  nebula seen in the radio continuum and optical images. The $EM$ profile shows a  sharp border towards the west of the nebula, while towards the east it decreases smoothly, indicating the existence of an electron density gradient. The observed $EM$ profile is in agreement with the results of \citet{c00}, who reported the existence of an electron density gradient in the east-west direction.  Using the peak values \hbox{$EM$ =  18000 pc cm$^{-6}$} (obtained at \hbox{{\it(l,b)} $\approx$ (289\fdg50, +0\fdg11)} for \ngc)  and \hbox{$EM$ = 7000 cm$^{-6}$}  (obtained at  \hbox{{\it(l,b)} $\approx$ (289\fdg53, +0\fdg18)} for MBO), and considering a pure hydrogen plasma with a depth  along the line of sight equal to the size in the plane of the sky, electron density estimates \hbox{$n_e$ = 75 $\pm$ 14 cm$^{-3}$} and \hbox{$n_e$ = 45   $\pm$ 9 cm$^{-3}$} are obtained for \ngc and MBO, respectively. 
\begin{figure*}
\centering
\includegraphics[width=500pt]{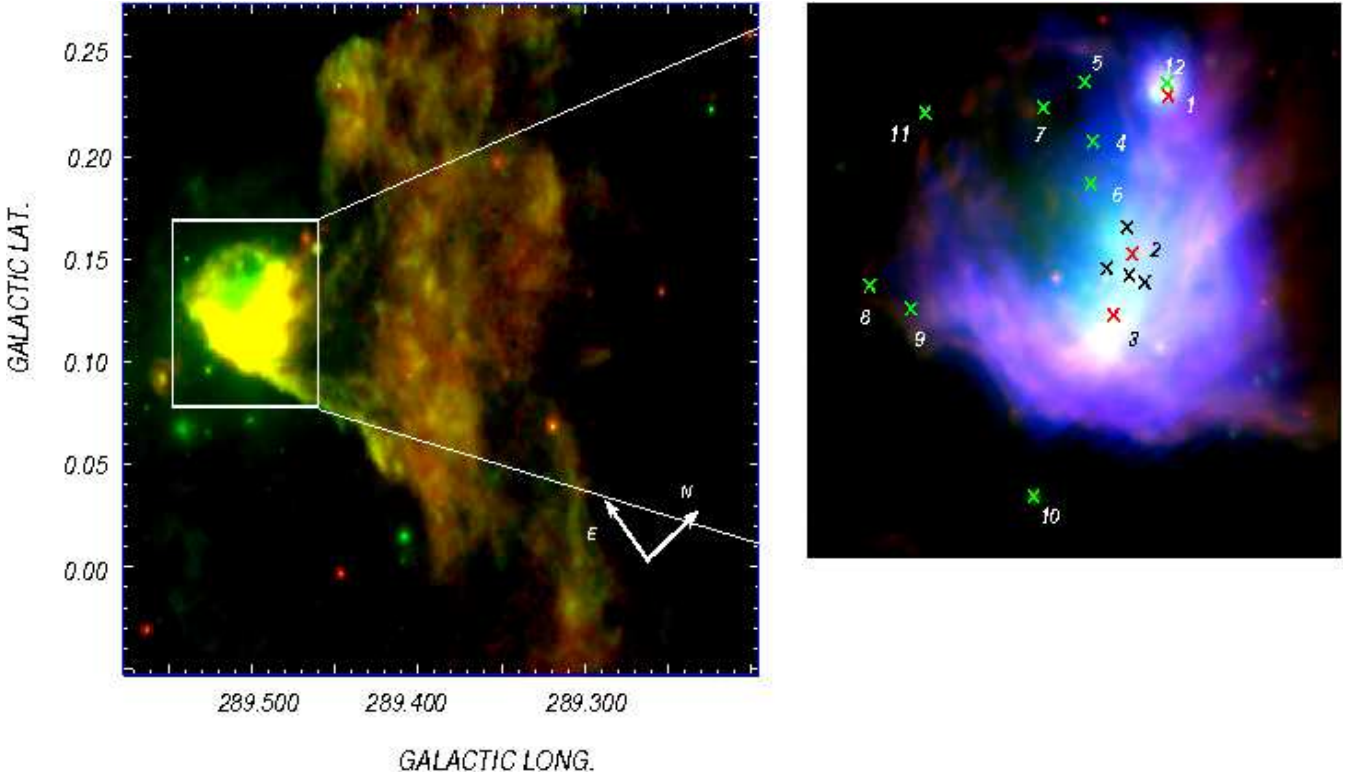}
\caption{{\it Left panel} Composite image of \ngc and its environs. Red and green show emission at 8.13 $\mu$m (MSX) and 24 $\mu$m (MIPSGAL). The color scale goes from 1$\times$10$^{-5}$  to 4$\times$10$^{-5}$ \hbox{W m$^{-2}$ sr$^{-1}$} and from 27 to 100 \hbox{MJy ster$^{-1}$}, respectively. The white rectangle encloses the IR counterpart of \ngc (IRK).    {\it Right panel}: Composite image of \ngc. Red, green and blue show emission at 8 $\mu$m (IRAC-GLIMPSE), 24 and 70 $\mu$m (MIPSGAL), respectively.   Colors scales range from 40 to 500 \hbox{MJy ster$^{-1}$} (24 $\mu$m), from 300 to 2000 \hbox{MJy ster$^{-1}$} (70 $\mu$m),  and  from 30 to 400 \hbox{MJy ster$^{-1}$}  (8 $\mu$m). The location of the members of Pis 17 is indicated with black crosses. MSX and 2MASS candidate YSOs  are indicated with red and green crosses (see text).}  
\label{fig:msx-halfa}
\end{figure*}

As a different approach   to \ngc, the rms electron density and ionized mass can be obtained  using the spherical model of \citet{mh67} and the flux density at 4800 MHz. Assuming a constant electron density,  and a radious \hbox{$R_{\rm HII}$ = 1.25 pc}, we obtain \hbox{$n_e$ = 54 $\pm$ 13 cm$^{-3}$} and \hbox{$M_{\rm HII}$ = 11 $\pm$ 3 \msun}. 
Thus, the electron density of \ngc is about a factor of 5 lower than the density of Component 1, which clearly indicates that the nebula has expanded. 
An estimate of the filling factor \hbox{($f$ = $\sqrt{n_e/n_e'}$)} can be obtained by taking into account the maximum electron density derived from optical lines \hbox{($n_e'$ = 154$^{+52}_{-45}$ cm$^{-3}$; \citealt{c00})}. We obtain $f$ =  0.5 - 0.8. Then, the ionized mass for $f$ = 0.5 - 0.8 is in the range 8 - 10 \msun.

Regarding MBO, rms electron densities and masses estimated from the image at 843 MHz are \hbox{$n_e$ $\simeq$ 33 \cm3} and \hbox{$M_{\rm HII}$ $\simeq$ 6 \msun}, respectively.


\subsection{Infrared Emission}

The emission distribution at 8.13 $\mu$m (MSX-A band) superposed to the image at 24 $\mu$m (MIPSGAL) is shown in the left panel of  Fig.~\ref{fig:msx-halfa}. The \msx Band {\rm A} includes the strong emission features at 7.7 and 8.6 $\mu$m attributed to {\rm PAH} molecules, which are considered tracers of {\rm UV}-irradiated {\it Photodissociated Regions} (PDR) \citep{ht97}. This image displays a small and strong feature, from hereonwards dubbed the {\it  IR Knot}, or {\rm IRK} for short (indicated with a white rectangle), which is  coincident with the location of \ngc, and a weaker and more extended  emission region detected in the north-western area of the image. The last feature will be referred to as the {\it Extended IR Emission}, or {\rm EIE} for short. IR emission at 8.3 and 24 $\mu$m appears mixed along the whole feature. A region of low  IR emission is seen between the IRK and the EIE. The bright rim of SFO 62 is detected in the IR as the bright filament observed from \hbox{($l, b$) $\simeq$ (289\fdg42, +0\fdg03)} to  \hbox{($l, b$) $\simeq$ (289\fdg50, +0\fdg08)},  which appears to be the southernmost boundary of EIE. The presence of PAH emission coincident with this bright rim  suggests the existence of a PDR at the southern edge of Component 1.

The right panel of Fig.~\ref{fig:msx-halfa} shows a detailed image of the IRK, at 8.0 $\mu$m ({\it Spitzer}-IRAC band 4, in red), 24 $\mu$m (MIPSGAL, in green), and 70 $\mu$m (MIPSGAL, in blue). The emissions   at 8 and 70 $\mu$m  show a bright half shell-like feature, which encloses the position of the stars of Pis 17, indicated as black crosses. Arc-shaped faint filaments are also detected in the 8 $\mu$m emission in the northeastern section of IRK.   The morphology of the IR emission is compatible    with the cometary-shape of the \hii region and very likely indicates the presence of a  PDR between \ngc and clump A. The  MIPSGAL  emission  at 24 $\mu$m, which is projected at the center of the structure,  indicates the existence of warm dust inside the H{\sc ii} region (as the cases of N10 and N21, \citealt{w08}).   
\begin{figure}
\centering
\includegraphics[width=250pt]{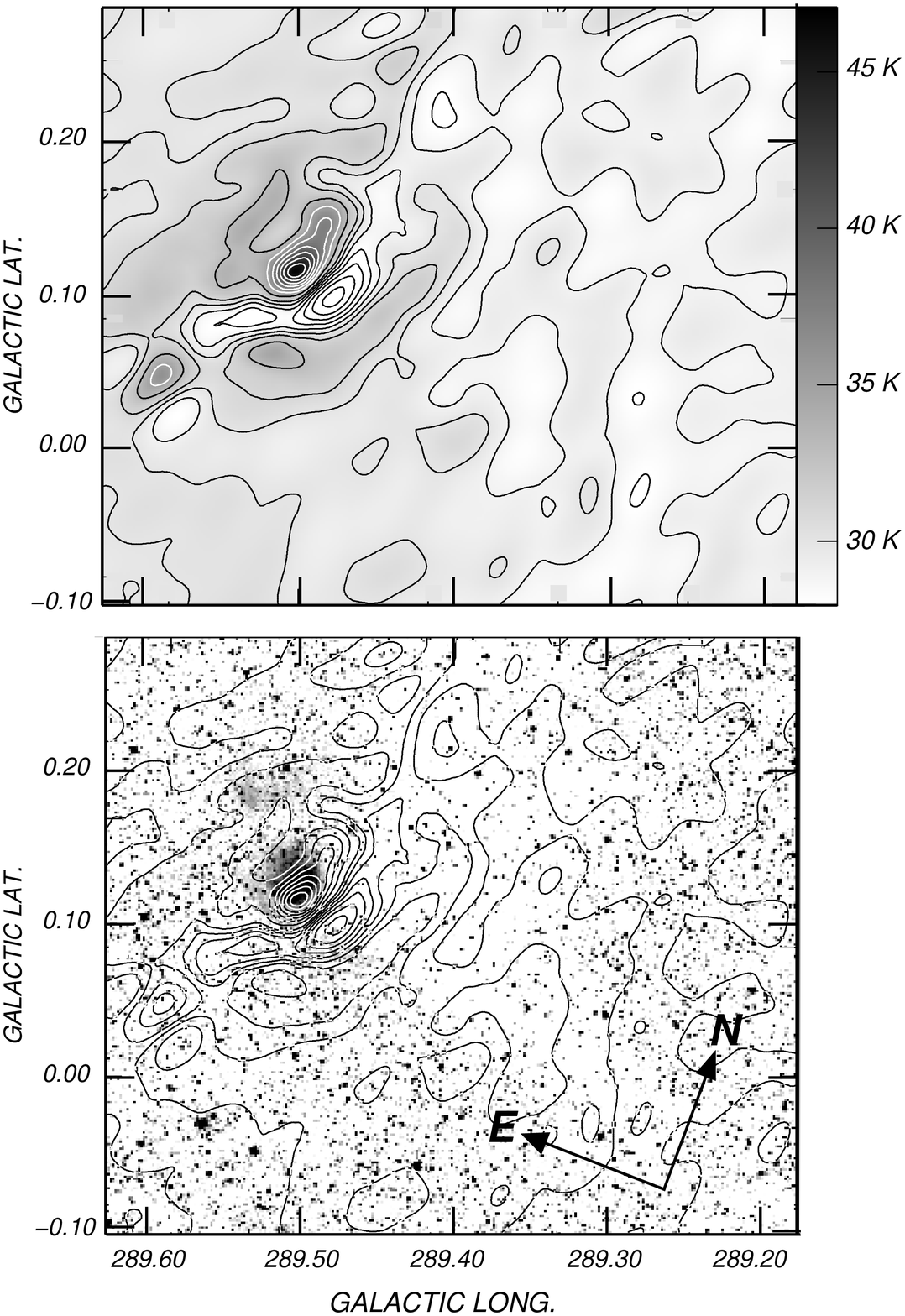}
\caption{ {\it Upper panel:} $T_{\rm d}$ distribution estimated from HIRES IRAS images at 60 and 100 $\mu$m. Contour levels go from 25  to 31 K in steps of 1 K, and from 33 K in steps of 2 K.   {\it Lower panel:} Overlay of the $T_{\rm d}$ distribution (contours) and the DSSR image.}
\label{fig:tdust}
\end{figure}

IRAS 60 and 100 $\mu$m  data show dust with color temperatures between about 20 to 100 K, which corresponds to the ``cool dust component'' (see \citealt{sr11}, and references therein). A dust temperature ($T_{\rm d}$) map was produced using the equation
\begin{equation}\label{td} 
\qquad T_{\rm d}\ =\ 95.94\ / \  \textrm{ln}(B_n)    \qquad ({\rm K})
\end{equation}
\citep{d90,wh92,ci01}, where $B_n$ = 1.667$^{(3+n)}$$\left(F_{100}/F_{60} \right)$ is the modified Planck function, with  $F_{100}$ and $F_{60}$ being the 100 $\mu$m and 60 $\mu$m fluxes, respectively. The parameter $n$ = 1.5  is related to the absorption efficiency of the dust ($k_{\nu}\ \propto\ \nu^n$, normalized to 40 cm$^2$ g$^{-1}$ at 100 $\mu$m). The obtained dust temperature map is shown in the upper panel of Fig. \ref{fig:tdust}.  Dust temperature goes from 25 K to 46 K. Fig. \ref{fig:tdust} (lower panel) also shows that  the  region with  the highest  dust  temperatures ($\sim$ 46 K) coincides with the brightest section of \ngc. This temperature is in good agreement with those obtained in RCW 121 and RCW 122 \citep{a08}  and in NGC 6357 \citep{cb11}  although they are slightly higher than  typical values for  H{\sc ii} regions \hbox{($\sim$ 30 K)}  (see \citealt{cn08,ci09,v10}). The stellar UV radiation field of the stars in NGC 3503 is responsible for the heating of the dust immersed in the \hii region. A noticeable feature is a fringe of cold ($\sim$ 25 K) dust surrounding the position of \ngc, which coincides with the region showing low emission at 8.3 $\mu$m placed between IRK and EIE  described above. An extended region of warmer ($\sim$ 36 K) dust is detected to the northeast of \ngc, coincident with the position of MBO, indicating that this optical  feature has a  radiatively heated dust component.
In Table \ref{table:IR} we summarized the IR and dust parameters  for IRK and EIE. The averaged dust temperatures ($\bar{T}_{\rm d}$) are calculated using Eq. \ref{td}   and flux densities at 60 and 100 $\mu$m obtained integrating several polygons around the sources.   Dust masses  are derived from
\begin{equation}\label{}
 M_d = m_n\  F_{60}\ d^2 \ (B_n^{2.5}\  -\ 1)                 \qquad ({\rm M}_{\odot}) 
\end{equation}
where {\it d} the distance in kpc, $F_{60}$ is  given Jy, and \hbox{$m_{1.5}$ = 0.3 $\times$ 10$^{-6}$}. Based on the molecular gas mass derived for Component 1, the ionized gas of \ngc and MBO, and the dust masses obtained for the IRK and the EIE ($\sim$ 10 \msun), we derive a mean weighted gas-to-dust ratio of about \hbox{$\sim$  800 $\pm$ 300}. This ratio is a factor of about $\sim$ 5  higher than the average values assumed for the Galaxy (140 $\pm$ 50; \citealp{t00}). Note however that dust with temperatures $T_d\ \lesssim\ 20$ K   (``cold dust component''), which is detected at wavelengths  not surveyed by IRAS \hbox{($\lambda$ $>$ 100 $\mu$m)},  dominates the dust mass by a factor larger than $\sim$ 70 with respect to the cool dust component \citep{sr11}.    This  notoriously decreases  the gas-to-dust ratio. The dust temperature estimated for EIE may suggest the existence of ionizing sources inside Component 1. However, a search for OB stars embedded in Component 1 performed using the available VIZIER catalogues failed to detect such sources.

 To investigate the presence of protostellar candidates possibly related to NGC\,3503, we used data from the MSX, 2\,MASS, and IRAS point source catalogs. We searched for point sources in a region of about 10\arcmin\ in size centered at the position of the \hii\ region. Taking into account sources with  flux quality $q >$ 2, a total of 69 MSX point sources were found projected onto the area.  Based on $F_{21}$/$F_{8}$  and $F_{14}$/$F_{12}$ ratios, where $F_8$, $F_{12}$, $F_{14}$, and $F_{21}$ are the fluxes at 8.3, 12, 14, and 21 $\mu$m, respectively, and  after applying the criteria summarized by  \citet{lu02}, we were left with three sources with $F_{21}$/$F_8 >$ 2 and $F_{14}$/$F_{12} <$ 1, which are classified as compact \hii regions (C\hii). This is indicative of stellar formation. The location of these sources is indicated with red crosses in Fig. \ref{fig:msx-halfa}.

\begin{table}
\caption{ Main infrared parameters inferred from IRAS fluxes at 60 $\mu$m and 100 $\mu$m}
\begin{center}
\begin{tabular}{lcc}
\hline      
  {\it Parameter}  & IRK & EIE \\
 
\hline\hline  
&&\\           

S$_{60}$ (Jy)    & 1470 $\pm$ 20          &  700 $\pm$ 15     \\  
S$_{100}$ (Jy)   & 2170 $\pm$ 25 &    1970 $\pm$ 25            \\
 $\bar{T}_d$  (K)   &      $\sim$ 37           & $\sim$ 28     \\ 
$M_d$   (\msun)      &  $\sim$ 3    & $\sim$ 7   \\
&&\\
\hline
\label{table:IR}
\end{tabular}
\end{center}
\end{table}

\begin{table*}
\begin{center}
\caption{Candidate YSOs obtained from the MSX and 2\,MASS catalogues.}
\label{yso}
\begin{tabular}{ccccccc}
\hline
 & ($l,b$) & MSX source & F$_8$ (Jy) & F$_{12}$ (Jy) & F$_{14}$ (Jy) & F$_{21}$ (Jy) \\
\hline
  &  &      &  &  &  &  \\
1 & 289.48,+0.14  &   G289.4859+00.1420& 1.49250 & 1.8145 & 0.9005 & 0.3180 \\
2 & 289.49,+0.12   & G289.4993+00.1231& 0.82699 & 1.4887 & 1.2237 & 4.3818 \\
3 & 289.50,+0.11   & G289.5051+00.1161& 0.87911 & 2.8632 & 1.6759 & 5.6330 \\
&&&&&&\\
\hline
 &  ($l,b$) & 2\,MASS source & $J$ (mag) & $H$ (mag) & $K$ (mag) & \\
\hline
  &      &  &  &  &  &  \\

4  & 289.49,+0.14    & 11011934-5949422& 13.277& 13.047& 12.806 &\\
5  & 289.50,+0.15   & 11012021-5949123& 15.084& 14.837& 14.506 &\\
6 & 289.50,+0.13   & 11011933-5950037& 14.539& 14.100& 13.696 &\\
7  & 289.50,+0.14  & 11012329-5949269& 14.862& 14.380& 14.027 &\\

8  & 289.54,+0.13    & 11013591-5951037& 15.237& 14.592& 14.106 &\\

9  & 289.53,+0.12   & 11013263-5951140& 15.633& 14.946& 14.404 &\\

10  & 289.52,+0.10  & 11012238-5952449& 12.747& 11.744& 10.836 &\\
11  &289.52,+0.15   & 11013238-5949339& 14.754& 14.243& 13.873 &\\
12  & 289.48,+0.14  & 11011390-5949099& 14.888& 13.661& 12.827 &\\

 &      &  &  &  &  &  \\
\hline 
\end{tabular}
\end{center}
\end{table*}

 Using the 2\,MASS catalog \citep{cu03}, which provides detections in $J$, $H$ and $K_s$ bands, we searched for point sources with infrared excess. Taking into account sources with signal-to-noise ratio (S/N) $>$ 10 (corresponding to quality ``AAA''), we found 5528 sources projected onto a circular region of 5\arcmin\ in radius.
 Following  \citet{co05}, we determined the parameter \hbox{$q$ = ($J$ - $H$) - 1.83 $\times$ ($H$ - $K_s$)}. Sources with $q <$ --0.15 are classified as objects with infrared excess , i.e. candidate YSOs. By  applying the above criteria, we found only 9 sources projected onto \ngc. The location of these sources is also indicated in Fig. \ref{fig:msx-halfa}  with green crosses. The location of the  2\,MASS source 11011390-5949099 (source $\#$12) is almost coincident with the MSX source G289.4859+00.1420 (source $\#$1). These sources are projected onto an intense  mid-IR clump located in  the northeastern border of the bright half shell-like feature,  at \hbox{($l,b$) = (289\fdg48,+0\fdg14 )} (see Fig. \ref{fig:msx-halfa}, right panel), and onto a small enhancement in the radio continuum emission detected at 4800 MHz (see Fig. \ref{fig:champ-ir-halfa} below).    These characteristics make this object an excellent candidate for investigating star formation with high angular resolution observations.  The names of the YSO candidates, their position, fluxes and magnitudes at different IR wavelengths are listed in Table \ref{yso}.


\section{Discussion}

\subsection{The ionized gas}

 The number of ionizing Lyman continuum photons ($N_{\rm   Lyc}$) needed to sustain the ionization in \ngc  can be  calculated using the equation given by \citet{sr90}
\begin{eqnarray}
\quad    N_{\rm Lyc}\ =\ 4.45 \times 10^{48}\ T_e^{-0.45}\ S_{(4800\ {\rm MHz})}\ d^2
\label{formula:nlym}
\end{eqnarray}
 where $d$ is the distance in kpc, and $S_{(4800\rm \ MHz)}$ is the flux density at 4800 MHz in Jy. We obtain  \hbox{$N_{\rm Lyc}$ = (1.8 $\pm$ 0.4) $\times$ 10$^{47}$ s$^{-1}$}. This number is a lower limit to the total number of Lyman continuum photons required to maintain the gas ionized, since about 25 - 50 $\%$ of the UV photons are absorbed by interstellar dust in the H{\sc ii} region \citep{i01}. Consequently, we need \hbox{$N_{\rm Lyc}$ $\approx$ 3.6 $\times$ 10 $^{47}$ s$^{-1}$}. The number of Lyman continuum photons emitted by a  B0 V star  is $N_{\rm Lyc}^*$ =  1.07 $\times$ 10$^{48}$ s$^{-1}$    \citep{st03},  which is capable  of ionizing \ngc, in agreement with previous results by \citet{T04} and   \citet{pco10}. The last authors  asserted one B0 V star and three B2 V stars, belonging to the open cluster Pis17, inside the nebula as ionizing  sources. Since  the ionizing photons emitted by a B2 V star are significantly fewer  than those of a B0 V star, their contribution to the energetics of the nebula can be neglected.

In Sec. 3.2  we pointed out on the discrepancy between the electron and molecular densities. The density of  Component 1 is about a factor of 5  higher than the electron density of \ngc, which clearly indicates that \ngc  has been expanding as a result of unbalanced pressure between ionised and molecular gas. 
In order to estimate the dynamical age ($t_{\rm dyn}$) of the \hii region  we used the model of \citet{dw97}. The radious of  an \hii region ($R_{\rm HII}$)  in a uniform medium is given by
\begin{equation}
\quad  \frac{R_{\rm HII}}{R_{\rm S}}\ =\ \left(1\ + \frac{7\ {\rm v}_{\rm s}\ {t}_{\rm dyn}}{4\ R_{\rm S}}\right)^{4/7} 
\end{equation}
were $R_S$ is the radious of the Str\"omgren sphere  \citep{s39} before expanding, given by \hbox{$R_{\rm S} = \left(3 N^*_{Lyc} / 4 \pi\ (2n_{\rm H_2})^2 \alpha_{\beta} \right)^{1/3}$}, and v$_{\rm s}$ is the sound speed in the ionized gas \hbox{($\sim$ 10 \hbox{\kms})}. For \hbox{$R_{\rm HII}$ $\simeq$ 1.25 pc} (1\farcm5 at a distance of 2.9 kpc),   an ambient density  \hbox{$n_{\rm H_2}$ = 400 $\pm$ 240  cm$^{-3}$} (see Section 3.1.3), and $N^*_{Lyc}$ =  1.07 $\times$ 10$^{48}$ s$^{-1}$, we infer \hbox{$t_{\rm dyn}$ $\approx$ 2 $\times$ 10$^5$} yr. The expansion velocity of the \hii region  can be estimated by means of   $\dot{R}_{\rm HII}$ = v$_s$ $\left( \frac{R_{\rm HII}}{R_s} \right)^{-3/4}$, yielding an  expansion  velocity  of about  $\sim$ 5  km s$^{-1}$.  The obtained dynamical age and expansion  velocity are in agreement with those obtained in typical \hii regions (Gum 31, \citealt{cn08}; Sh2-173, \citealt{ci09}).

 The total number of ionizing Lyman continuum photons needed to sustain the ionization in MBO  is \hbox{$N_{\rm Lyc}$ $\approx$ 2.5 $\times$ 10$^{47}$ seg$^{-1}$}.  To search for stars    that can provide the necessary  UV photons, we  used the available VIZIER catalogues. CP-59 2951 is a B1 V star placed at \hbox{{\it(l,b)} = (289\fdg511, +0\fdg195)} \citep{bu99}. Using the catalogued spectral type,  the visual magnitude, and the calibration of \citet{sk82}, we estimated a distance \hbox{$d$ = 2.7 $\pm$ 0.8 kpc}. The continuum ionizing photons emitted by the star  are  \hbox{$N_{\rm Lyc}^*$ = 3.2 $\times$ 10$^{46}$ s$^{-1}$} \citep{sm02}. Although the distance of this star is in agreement (within errors) with the distance of \ngc, the value of $N^*_{\rm Lyc}$ is almost one order of magnitude lower than the required to ionize MBO. The presence of Pis17 at a projected distance of  3.4 pc ($\sim$ 4$'$ at a distance of 2.9 kpc) indicates that the contribution of the star cluster to the ionization of MBO can not be ruled out.

\subsection{Star formation process}

 The presence  of 12 candidate YSOs projected onto the  \ngc-region suggests that they may have been triggered by the expansion of the \hii region through the ``collect and collapse'' model, which  indicates that expanding nebulae compress gas between the ionization and the shock fronts, leading to the formation of  molecular cores where new stars can be embedded. Using the analytical model of \citet{wi94} for the case of expanding \hii regions, we derived the time when the  fragmentation may have occurred ($t_{frag}$), and  the size of the \hii region at $t_{frag}$ ($R_{frag}$), which are given by
\begin{eqnarray}
t_{frag} [10^6 yr]\ =\ 1.56 a_2^{4/11}\ n_3^{-6/11}\ N_{49}^{-1/11}
\end{eqnarray}
\begin{eqnarray}
R_{frag} [pc]\ =\ 5.8 a_2^{4/11}\ n_3^{-6/11}\ N_{49}^{1/11}
\end{eqnarray}
 where $a_2$ is the sound velocity in units of 0.2 \kms,  $n_3 \equiv  n_{H_2}/1000$, and $N_{49}\equiv N_{Lyc}^*/10^{49}$. Adopting for this region 0.3 \kms for the sound velocity, which corresponds to temperatures of 10-15 K in the surrounding molecular clouds (see Section 3.1.2), we obtained \hbox{$t_{frag} \sim$ 3.5 $\times$ 10$^6$  yr}, and  \hbox{$R_{frag} \sim$ 7.5 pc}. Considering that the values of $t_{frag}$ and $R_{frag}$ are larger than $t_{dyn}$ and $R_{HII}$ (see Section 4.1), we can conclude that the fragmentation  in the edge of NGC\,3503 is doubtful.

\subsection{Kinematics of clump A}

In section 3.2  we reported the existence of a velocity gradient across clump A.  Velocity gradients in molecular cores/clumps were usually interpreted as gravitationally bound rotation motions (particulary in dense and small molecular cores). Several  theoretical models predict cloud flattening perpendicular to the rotation axis in response to centrifugal stress, which at first glance seems to be suitable for clump A. 
In order to investigate the dynamical stability of clump A, we use the parameter $\beta$ defined by \citet{go93}  to quantify  the dynamical role of rotation  by  comparing the rotational kinetic energy to the gravitational energy. Thus, $\beta$ can be written  as
\begin{equation}\label{eq:Tr}
\qquad    \beta = \frac{(1/2)\ I\ \omega'^2}{q\ G\ M^2/R}\ =\ (1/2)\ (p/q)\ \frac{\omega'^2\ R^3}{G\ M}
\end{equation}
where $I$ is the moment of inertia ($I=pMR^2$), $q G M^2/R$ is the gravitational potential energy, and $\omega'=\omega/sin(i)$, where $i$ is the inclination of the cloud along the line of sight. Considering \hbox{$p/q$=0.22} (see \citealt{go93}), \hbox{sin($i$) = 1}, \hbox{$\omega$=0.15 \kms pc$^{-1}$} (see Sect 3.1.2), \hbox{$R$ $\sim$ 4.5 pc}, and a lower limit mass to clump A  \hbox{($M$ $\sim$ 3 $\times$ 10$^{3}$ \msun)},  we obtain \hbox{$\beta\ \approx$ 0.02}. This extremely low value of $\beta$ indicates that the effect of rotation, if exists, is not significant in mantaining the dynamical stability of \hbox{clump A}.  This might weaken the rotating cloud interpretation. Furthermore, Figs.  \ref{fig:pos-vel} and  \ref{fig:mosaico1} show that  only molecular gas at more negative velocities ($\sim$ $-$27 \hbox{\kms} / $-$26 \hbox{\kms})  coincide with \ngc, which suggests that the velocity gradient might be a direct consequence of an interaction between clump A and \ngc.

  Therefore, although rotation can not be entirely ruled out with the present data,   we consider different origins for the velocity gradient observed across clump A, namely, 1)  the expansion of the nebula at $\sim$ 5 \hbox{\kms} (see Sect 4.1)   has been accumulating molecular gas   behind the shock front which originated the expansion of clump A at approximately the same velocity of the ionized gas, as expected according to the models of \citet{hi96}, 2) the velocity gradient of clump A is the consequence of a collision between Component 1 and another molecular cloud, which in turn, might have induced the formation of Pis 17 and the candidate YSOs reported in Sect. 3.3. Although they are rare, cloud-cloud collisions can lead to gravitational instabilities in the dense, shocked gas, resulting in triggered star formation (see \citealt{elm98}, and references therein), 3) clump A is actually composed by different subclumps at different velocities, which are not resolved by the NANTEN observations.  In this case, \ngc  might be related with a subclump at more negative velocities. 

 Further high-resolution studies with instruments like APEX may help to clarify this question.

\subsection{The PDR at the interface between NGC 3503 and clump A}

 As mentioned in Sect 3.1.2, clump A exhibits an excitation temperature  \hbox{$T_{exc}$ $\geq$ 17.7 K}, which is also the highest excitation temperature along Component 1. This temperature is  higher than expected inside molecular cores if only cosmic ray ionization is considered as the main heating source \hbox{($T$ $\sim$ 8 - 10 K,   \citealt{vdt00})}, which implies that additional heating processes are present close to clump A. Very likely, clump A is being externally heated through the photoionisation of its surface layers \citep{U09}  as a consequence of its proximity to \ngc. This scenario is in line with  the presence of the PDR at the interface between \ngc and clump A (see \hbox{Sect. 3.3}). In this context, it would be instructive to make a simple comparison of the PDR dust surface temperature, with the dust temperature obtained before    towards the edge of \ngc (see Fig.\ref{fig:tdust}).

The structure of a PDR is governed by the intensity of the UV radiation field  impinging onto the cloud surface ($G_0$) in units of the \citet{ha68}   FUV flux (1.6 \x 10$^{-3}$ ergs cm$^{-2}$ s$^{-1}$), with $G_0\propto  N_{\rm Lyc}^{-2/3}\ \  n_e^{4/3}\ \ L_{\star}\ \chi$ \citep{ti05}, where $\chi$ is the fraction luminosity over 6 eV,  and $L_{\star}$ is the luminosity of the star. Adopting $N_{\rm Ly}$ = 1.07$\times$10$^{48}$ s$^{-1}$ and  $L$ = 7.6 $\times$ 10$^4$ $L_{\odot}$ \citep{st03}, $\chi$ = 1, and considering $n_e\simeq$ 75 -- 154 \cm3 (see Sect. 3.2), a range of $G_0\simeq$ (0.5 - 1.5) $\times$ 10$^3$ is obtained. The distribution of dust temperature in the PDR  ($T_d^{pdr}$) is governed by the absorption of the stellar photons which are reemited by dust grains as IR photons. Following \citet{ti05}, we can relate $T_d^{pdr}$ with the visual absorption along the PDR ($A_{\rm v}$) as
\begin{equation}\label{g0}
(T_d^{pdr})^5 \simeq \nu_0\ G_0\ e^{(-1.8A_{\rm v})} +\ \ln(3.5\times10^{-2}\tau_{100}\ T_0)\ \tau_{100}\ {T_0}^6
\end{equation}
\noindent{where $\nu_0$ is the frequency at 0.1  $\mu$m, $\tau_{100}=10^{-3}$ is the effective optical depth at 100 $\mu$m, and $T_0$ is the temperature of the slab estimated as \hbox{$T_0$ = 12.2 {$G_0$}$^{1/5}$}}. Considering both,  the position of NGC\,3503 at the edge of clump A, and  that the low value of visual absorption derived for the members of Pis\,17 ($\bar{A}_{\rm v}$ = 1.6 mag,  \citealt{pco10}) is  probably due to the interstellar absorption in the line of sight of the \hii region, we can assume $A_{\rm v}\sim$ 0 for the surface of the PDR. Taking into account the range of $G_0$ and Eq.~\ref{g0}, we obtain \hbox{$T_d^{pdr}$ $\simeq$ 40 - 55 K}, in accordance with the observed dust color temperature estimated for    NGC\,3503 \hbox{($\sim$46 K)}.  This gives additional support to the existence of a PDR between \ngc and clump A.

\begin{figure*}
\centering
\includegraphics[width=170mm]{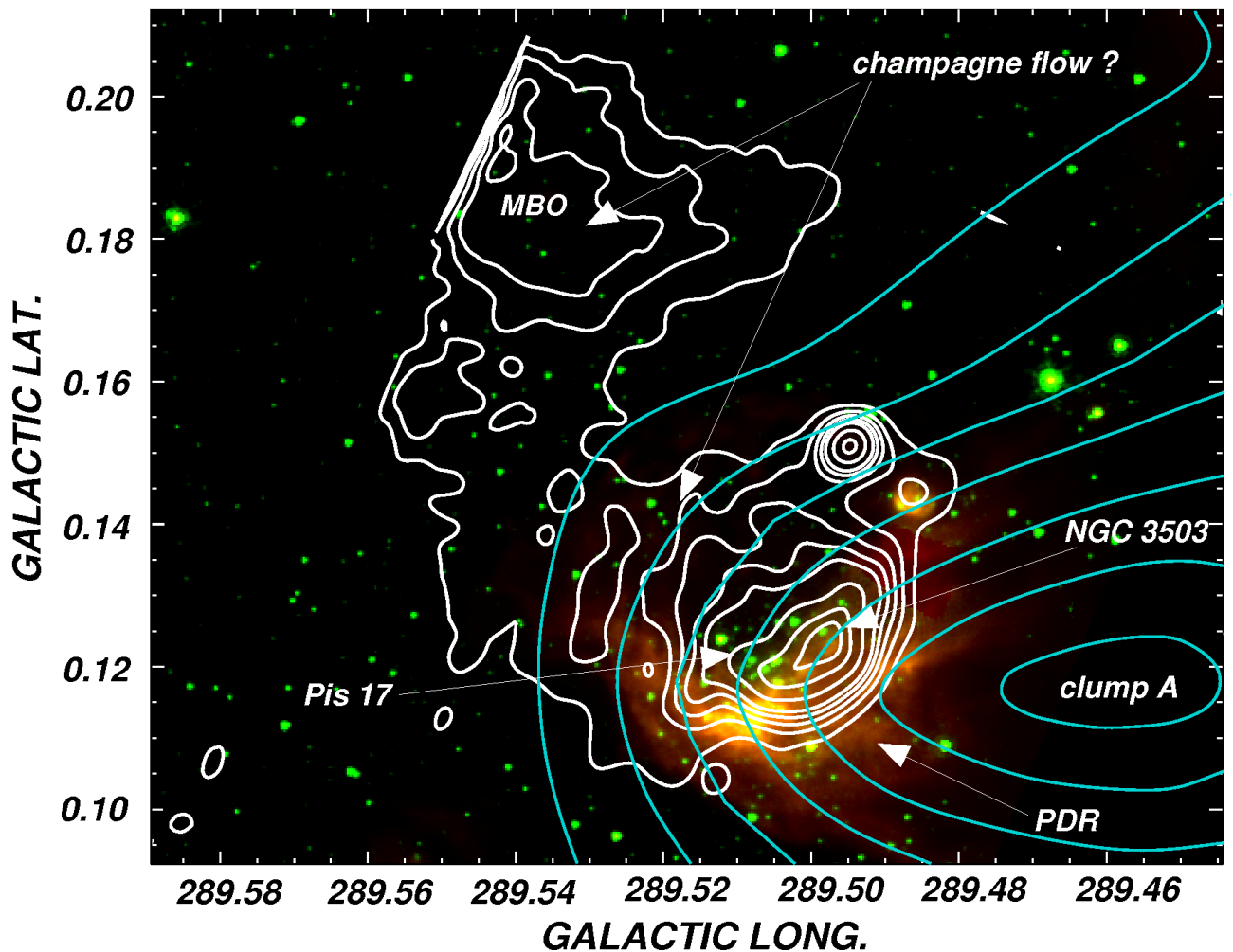}
\caption{ Composite image of \ngc and its environs. Red and green show emission at 8 and 4.5 $\mu$m (IRAC-GLIMPSE), respectively. White and light blue  contours show the radiocontinuum 4800 MHz and CO line emission, respectively.      }
\label{fig:champ-ir-halfa}
\end{figure*}

\subsection{Possible scenario}

Figure \ref{fig:champ-ir-halfa}   displays a composite image of NGC3503 and its environs. White and light blue contours show the radio continuum emission at 4800 MHz corresponding to NGC3503 and MBO, and the CO emission, respectively. The color scale  shows the emission at 8 $\mu$m (red) and 4.5 $\mu$m (green).

As described in Sect. 3.3 and 4.3, emission attributed to PAHs encircles the southern and western borders of the \hii region, indicating the location of the PDR. The position of NGC 3503 near the border of clump A and the existence of an electron density gradient along the symmetry axis of the \hii region (with the higher electron densities close to the strongest CO emission) suggest  that the ionised gas is being  streamed away from the densest part of the molecular cloud. This is indicative that \ngc  is a blister-type \hii region \citep{i78}   that probably has undergone a champagne phase.     

The so-called {\it Champagne flow model} \citep{tt79,btt81,ttb82}    proposes that an expanding \hii region placed at the edge of a molecular cloud eventually reaches the border of the cloud and expands freely in the lower density surrounding gas, originating an extended  \hii region with a characteristic density distribution. The \hii region becomes density bounded towards the lower density region, while it is ionization bounded towards the molecular cloud. This scenario was formerly proposed by \citet{c00}  to explain the electron density gradient observed across NGC 3503. The location of Pis 17 close to  the brightest radio continuum region is  consistent with a projection effect \citep{ytb83}. In this scenario, Pis 17 originated NGC 3503, which  reached the northeastern border of \hbox{clump A} after 2$\times$ 10$^5$ yr. This time  represents a lower limit to the age of the \hii region, since the inferred main-sequence life time of the main ionizing star is about \hbox{(3 -5) $\times$ 10$^6$ yr} \citep{m98}. The leakage of ionized gas  and  UV photons might have contributed to the formation of MBO, since its location (along the symmetry axis of \ngc) and its low density (about half of \ngc) suggest that this feature may consist of ionized gas which has scaped from \ngc after a time of  2$\times$ 10$^5$ yr.

The velocities of both the molecular and ionized gas are compatible with the champagne scenario. The mean velocity of Component 1 (--24.7 \hbox{\kms}, see Sect. 3.1.1) corresponds to the natal cloud where  NGC 3503 originated. Molecular gas having more negative velocities is mainly linked to clump A and may represent material moving away from the ionized gas placed in front of the \hii region  (as a result of expanding motions or an external impact  over Component 1). The ionized gas, having a velocity of --21 \hbox{\hbox{\kms}} \citep{g00}, might be receding the observer.  We note, however, that the Fabry-Perot observations by \citet{g00}   have a spectral resolution of 5 \hbox{\kms}.

\ngc resembles the two well studied cases of S305 and S307. These \hii regions show a non spherical  morphology in the 1465 MHz map of \citet{fi93}, and they are placed close to $^{12}$CO emission peaks, which suggests that they are located close to the hottest part of their parental molecular clouds (see \citealt{ru95}, and references therein). Both objects show  significant electron density dependence on position, which can also be interpreted as radial gradients \citep{c00}. Furthermore, both \hii regions show  discrepancies of about \hbox{$\sim$ 12 \hbox{\kms}} between the velocities of the ionized and molecular gas, probably due to a champagne effect producing a flow toward the observer \citep{ru95}. In the case of S307, the champagne scenario is reinforced by its half-shell shape, and the high density  in the brightest part of the nebula \citep{fh81,as86}

According to the champagne scenario, a velocity gradient along the symmetry axis of the nebula can be expected \citep{glg94}. In the case of \ngc increasing velocities with the distance to the molecular cloud are expected. Thus, high spectral resolution radio recombination are required to analyze the ionized gas velocities and may give additional support to our interpretation.

\section{Summary}

NGC\,3503 is a bright \hii\ region of $\sim$ 3\arcmin\ in size centered at \hbox{{\it(l,b)} = (289\fdg51, +0\fdg12)} located at a distance of  2.9 $\pm$ 0.4 kpc. The ionizing sources are B-stars belonging to the open cluster Pis\,17.  

With the aim of investigating the molecular gas and dust distribution  in the environs of the \hii\ region and analyzing the interaction of the nebula and Pis\,17 with their molecular environment, we analyzed $^{12}$CO(1-0) data of a region of 0\fdg 7 in size obtained with the NANTEN telescope with an angular resolution of 2\farcm 7,  radio continuum data of NGC\,3503 at 4800 and 8640 MHz obtained with the ATCA telescope (with synthesized beams of 21\arcsec\ and 13\arcsec, respectively), and data  at 843 MHz retrieved from SUMSS, and available IRAS, MSX, IRAC-GLIMPSE, and MIPSGAL images. 

The analysis of the CO data allowed the molecular gas linked to the nebula  to be mapped. This molecular gas (Component 1) has a mean velocity of --24.7 \hbox{\kms}, a total mass and density of (7.6 $\pm$ 2.1) $\times$  10$^3$ \msun and 400 $\pm$ 240 cm$^{-3}$, respectively,   and displays  two  clumps centered at \hbox{{\it(l,b)} = (289\fdg47, +0\fdg12)} (clump A) and  \hbox{{\it(l,b)} = (289\fdg32, +0\fdg03)}    (clump B). The morphological correspondence of the molecular emission with a large patch of high optical absorption adjacent to  NGC\,3503 is excellent. NGC\,3503 is projected near the border of clump A, with the strongest molecular emission adjacent to the highest electron density regions. The agreement of the molecular velocities with the velocity of the ionizad gas (--21 \hbox{\kms}) as well as the morphological correspondence with the nebula indicate that clump A is associated with NGC\,3503. The more negative velocities of the gas in clump A, coincident with the \hii\ region, are probably due to the expansion of the \hii\ region  or an external impact over Component 1.

The analysis of the radio continuum images confirms the electron density gradient previously found by \citet{c00}. The images show that NGC\,3503 exhibits a cometary morphology, with the higher density region  near the maximum of clump A. The rms electron density and the ionized mass amount to 54 $\pm$ 13 cm$^{-3}$ and 9 $\pm$ 3 \msun, respectively. A low density ionized region (MBO) located close to the lower electron density area of NGC\,3503 is also identified in these images.

Strong emission at 8 $\mu$m surrounds the bright radio continuum region of the nebula, indicating the presence of a photodissociated region at the interface between the ionized region and clump A. MIPSGAL emission at 24 $\mu$m shows the existence of warm dust inside the \hii\ region. Based on high resolution IRAS images at 60 and 100 $\mu$m, a mean dust color temperature and dust mass of 37 K and 3 \msun were estimated for \ngc.  The presence of candidate YSOs projected onto the HII region, detected using MSX and 2MASS point source catalogues, suggests the existence of protostellar objects in the neighbourhood of    NGC 3503, although there is no clear evidence of triggered star formation.

The location of NGC\,3503 at the edge of a molecular clump and  the electron density gradient in the \hii\ region suggests that NGC\,3503 is a blister-type \hii\ region that has undergone a champagne phase. In this scenario, the massive stars of Pis\,17 created NGC\,3503, which reached the border of the molecular cloud after 2 $\times$ 10$^5$ yr. The leakage of ionized gas and UV photons have probably contributed in the formation of MBO. Thus, the spatial distribution of the molecular gas and PAHs give additional support to  a scenario first proposed by \citet{c00}. The proposed scenario for \ngc may explain the slight difference between the main velocity of its parental molecular cloud \hbox{(--24.7 \hbox{\kms})} and the velocity of its ionized gas \hbox{(--21 \hbox{\kms})}.     High resolution radio recombination line observations may help to confirm the proposed scenario.

.

\begin{acknowledgements}

 We especially thank  Dr. James S. Urquhart and Dr. Mark A. Thompson for making  their unpublished radio continuum images  at 4800 MHz and 8640 MHz available to us.  We acknowledge the anonymous referees for their helpful comments that improved the presentation of this paper.  This project was partially financed by the Consejo Nacional de Investigaciones Cient\'ificas y T\'ecnicas (CONICET) of Argentina
under projects  \hbox{PIP 112-200801-02488} and \hbox{PIP 112-200801-01299}, Universidad
Nacional de La Plata (UNLP) under project \hbox{11G/093}, and
Agencia Nacional de Promoci\'on Cient\'ica y Tecnol\'ogica
(ANPCYT) under projects \hbox{PICT 2007-00902} and \hbox{PICT 14018/03}. 

This research has made use of the VIZIER database, operated at CDS, Strasbourg, France.
We greatly appreciate the hospitality of all staff members of Las Campanas Observatory of the Carnegie Institute of Washington. We thank all members of the NANTEN staff, in particular Prof. Yasuo Fukui, Dr. Toshikazu Onishi, Dr. Akira Mizuno, and students Y. Moriguchi, H. Saito, and S Sakamoto. We also would like to thank Dr. D. Miniti (Pont\'{\i}fica Universidad Cat\'olica, Chile) and Mr. F Bareilles (IAR) for their involment in early stages of this project

\end{acknowledgements}

\bibliographystyle{aa}
\bibliography{bibliografia-ngc3503}
 
\IfFileExists{\jobname.bbl}{}
{\typeout{}
\typeout{****************************************************}
\typeout{****************************************************}
\typeout{** Please run "bibtex \jobname" to optain}
\typeout{** the bibliography and then re-run LaTeX}
\typeout{** twice to fix the references!}
\typeout{****************************************************}
\typeout{****************************************************}
\typeout{}

}

\end{document}